\RequirePackage{fix-cm}

\documentclass[twocolumn]{svjour3}          
\usepackage[a4paper, total={7.42in, 8in}]{geometry}

\smartqed  

\usepackage{multicol}

\usepackage{graphicx}
\usepackage{multicol,lipsum}
\usepackage{amsmath}
\usepackage{dcolumn} 
\usepackage{enumitem}
\usepackage{graphicx}
\usepackage{dcolumn}
\usepackage{array,multirow}
\usepackage{bm}
\usepackage{amsmath}
\usepackage{epstopdf}
\usepackage{amsfonts}
\usepackage{amssymb}
\usepackage{mathrsfs}
\usepackage{epsfig}
\usepackage{tabularx}
\usepackage[dvipsnames]{xcolor}
\RequirePackage[colorlinks,citecolor=blue,urlcolor=blue,linkcolor=blue]{hyperref}
\usepackage{cite}
\newcommand{\be}{\begin{equation}}
	\newcommand{\ee}{\end{equation}}
\newcommand{\beq}{\begin{eqnarray}}
	\newcommand{\eeq}{\end{eqnarray}}
	\providecommand{\ed}{\mathrm{d}}

\newcommand{\arcsinh}{\mathrm{arcsinh}}
\newcommand{\arccosh}{\mathrm{arccosh}}

\begin{document}

\title{Motion of massive particles around a charged Weyl black hole and the geodetic precession of orbiting gyroscopes}

\author{Mohsen Fathi \and  Mona Kariminezhaddahka \and Marco Olivares \and
        J.R. Villanueva 
}

\institute{Mohsen Fathi \at  Instituto de F\'isica y Astronom\'ia, Universidad de Valpara\'iso, Avenida Gran Breta\~na 1111, Valpara\'iso, Chile\\ \email{mohsen.fathi@postgrado.uv.cl}
\and Mona Kariminezhaddahka \at 
Instituto de F\'isica y Astronom\'ia, Universidad de Valpara\'iso, Avenida Gran Breta\~na 1111, Valpara\'iso, Chile\\
               \email{mona.kariminezhad@gmail.com}
	\and Marco Olivares \at
              Facultad de
              Ingenier\'{i}a y Ciencias, Universidad Diego Portales, Avenida Ej\'{e}rcito
              Libertador 441, Casilla 298-V, Santiago, Chile\\
              \email{marco.olivaresr@mail.udp.cl}           
           \and
           J.R. Villanueva \at
               Instituto de F\'isica y Astronom\'ia, Universidad de Valpara\'iso, Avenida Gran Breta\~na 1111, Valpara\'iso, Chile\\
               \email{jose.villanueva@uv.cl}
}

\date{Received: date / Accepted: date}

\maketitle

\begin{abstract}
{The advanced state of cosmological observations constantly tests the alternative theories of gravity that originate from Einstein's theory. However, this is not restricted to modifications to general relativity. In this sense, we work in the context of Weyl's theory, more specifically, on a particular black hole solution for a charged massive source, which is confronted with the classical test of the geodetic precession, to obtain information about the parameters associated with this theory. To fully assess this spacetime, the complete geodesic structure for massive test particles is presented.}

\end{abstract}

\section{Introduction}

Classical physics description of falling particles in gravitational fields, has formed the foundations of general relativity. In this sense, the famous precessions in planetary orbits were described in the context of geodesic motion of falling massive particles in the gravitational fields produced by a central mass \cite{Misner:1973,Wald:1984}. Ever since the advent of general relativity and its success in responding to solar system tests, a strong attention to the investigation of the motion of astrophysical objects in gravitating systems,
like stars spiraling into black holes has been developed. Such theme, i.e. mass and its motion in general relativity has been also extended to other theories of gravity. For compact bodies,  the methods in this field of research also cover post-Newtonian frameworks and are applied for example to spiraling compact binaries, and even the self-force effects have found their way into the analysis of motion (for a very good review see Ref.~\cite{Blanchet:2011}).

Although the general relativistic results have shown to be in very good compatibility with observations of the gravitational waves \cite{Abbott:2016blz,Abbott:2016nmj,Abbott:2017vtc} and black hole optical appearance \cite{Akiyama:2019}, the appeal to alternative theories of gravity has, on the other hand reasonable sake. In fact, the late 1990s dark matter and dark energy scenarios, as the most mysterious problems to the modern cosmology, are supported by the observation of the flat galactic rotation curves \cite{Rubin1980}, the unexpected  gravitational lensing \cite{Massey2010} and the accelerated expansion of the universe \cite{Riess:1998,Perlmutter:1999,Astier:2012}. The tremendously weak interactions of dark matter with baryonic matter, and the impossibility of the detection of dark energy, have made some scientists to propose that the dark matter/dark energy scenarios stem from the incomplete knowledge that general relativity gives us about the behavior of the gravitational field. It is argued that, by adding particular components to, or changing the Einstein-Hilbert action, we can describe the cosmological anomalies by means of the resultant alternative gravitational theories (see Ref.~\cite{Clifton2012pr} for a review), without the necessity of the inclusion of the dark energy and dark matter.  

In the same effort, in 1980s, a relatively old theory, the Weyl conformal gravity which had been formulated by H. Weyl in 1918 \cite{Weyl1918mz}, was revived by Riegert \cite{Riegert1984}. This theory was then given an exact spherically symmetric static vacuum solution by Mannheim and Kazanas \cite{Mannheim:1989}. There, the authors showed that the problem of flat galactic rotation curves, could be avoided by calculating the radial velocities in the spacetime described by their solution. Therefore, beside Milgrom's post Newtonian dynamics (MOND) \cite{Milgrom:1983} which had been formulated in the same decade, Weyl conformal gravity was also proposed as an alternative to dark matter. The theory, as well, is intended to cover the dark energy related phenomena  \cite{Mannheim:2005,Nesbet:2012}. According to these interesting features, since the advent of the Mannheim-Kazanas solution, Weyl conformal gravity has been studied from several viewpoints \cite{Knox:1993fj,Edery:1997hu,Klemm:1998kf,Edery:2001at,Pireaux:2004id,Pireaux:2004xb,Diaferio:2008gh,Sultana:2010zz,Diaferio:2011kc,Mannheim:2011is,Tanhayi:2011dh,Said:2012xt,Lu:2012xu,Villanueva:2013Weyl,Mohseni:2016ylo,Horne:2016ajh,Lim:2016lqv,Varieschi2010,Hooft2010a,Hooft2010b,Hooft2011,Varieschi2012isrn,Varieschi2014gerg,Vega2014,Varieschi2014galaxies,Hooft2015,Deliduman:2015,Xu:2019,Turner:2020}. 

In this paper, we also consider Weyl conformal gravity to study the behavior of geodesic motion of massive particles near a static charged black hole introduced in Ref.~\cite{Payandeh:2012mj}. Recently, this black hole has undergone some classical tests in the context of light propagation in its exterior geometry \cite{Fathi:2020}. Here, we set the same spacetime as the background, to figure out the orbits of massive particles as they approach the black hole. In this regard, we can perform more classical tests on the black hole, in accordance with those tests done in the early times of general relativity. 

The paper is in fact divided into two main segments; the geodesic motions and a classical relativistic test. We organize our discussion as follows: For the first part of the paper, in Sec.~\ref{sec:solution}, we briefly introduce the Weyl field equations and its vacuum solution and ramify the black hole spacetime that we intend to study. In Sec.~\ref{sec:TimeLikeStructure} we construct a Lagrangian formalism to have a framework in studying the resultant effective potential of the black hole and its implied time-like trajectories. Several types of orbits, including the captures, scattered and critical trajectories are investigated in Sec.~\ref{sec:angular}. In Sec.~\ref{sec:radial} same methods are used to discuss other kinds of scattering and critical motions. In this section we also provide insights into the relative behaviors of the coordinate and proper time. For the second part of the paper and in Sec.~\ref{sec:precession}, we talk about the so-called geodetic effect imposed on the spin vector of an orbiting gyroscope, by considering a rotating frame on the background and compare our results with those inferred from general relativity. We conclude in Sec.~\ref{sec:conclusion}. In this paper, we work in geometric units, according to which, the speed of light and the Newton's gravitational constant are set to unity (i.e. $G=c=1$). Further discussions and related explanations will be given in appropriate places.

\section{The black hole solution}\label{sec:solution}
The conformal Weyl theory of gravity is described by the action
\begin{equation}\label{eq:IWeyl}
    I_W=-\mathcal{K}\int{\ed^4x\sqrt{-g}\,\,C_{\mu\nu\rho\lambda}C^{\mu\nu\rho\lambda}},
 \end{equation}
where $g=\mathrm{det}(g_{\mu\nu})$, and
\begin{multline}\label{eq:C}
C_{\mu\nu\lambda\rho} = R_{\mu\nu\lambda\rho}-\frac{1}{2}\left(g_{\mu\lambda}
R_{\nu\rho}-g_{\mu\rho}R_{\nu\lambda}-g_{\nu\lambda}R_{\mu\rho}+g_{\nu\rho}R_{\mu\lambda}\right)\\
+\frac{R}{6}\left(g_{\mu\lambda}g_{\nu\rho}-g_{\mu\rho}g_{\nu\lambda}\right)
\end{multline}
is the Weyl conformal tensor and $\mathcal K$ is a coupling  constant. The action $I_W$ is unchanged under the conformal transformation $g_{\mu\nu}(x) = e^{2 \alpha(x)} g_{\mu\nu}(x)$, in which $2 \alpha(x)$ is the local spacetime stretching. Combining Eqs.~\eqref{eq:IWeyl} and \eqref{eq:C}, we have
\begin{equation}\label{eq:IWeyl-2}
    I_W=-\mathcal{K}\int
\ed^4x
\sqrt{-g}\,\left(R^{\mu\nu\rho\lambda}R_{\mu\nu\rho\lambda}-2R^{\mu\nu}R_{\mu\nu}+\frac{1}{3}R^2\right).
\end{equation}
The Gauss-Bonnet term $\sqrt{-g}\,(R^{\mu\nu\rho\lambda}R_{\mu\nu\rho\lambda}-4R^{\mu\nu}R_{\mu\nu}+R^2)$
is a total divergence and does not contribute
to the equation of motion. The simplified action is therefore written as \cite{Mannheim:1989,Kazanas:1991}
\begin{equation}\label{eq:IWeyl-3}
    I_{W}=-2\mathcal{K}\int{\textmd{d}^4x}\sqrt{-g}\,\,\left(R^{\alpha\beta}R_{\alpha\beta}-\frac{1}{3}R^2\right).
\end{equation}
Applying  $\frac{\delta{I_W}}{\delta{g_{\alpha\beta}}} = 0$, leads to the Bach equation $W_{\alpha\beta} = 0$, with the Bach tensor defined as
\begin{eqnarray}\label{eq:Bach}
W_{\alpha\beta}&=&\nabla^\sigma\nabla_\alpha R_{\beta\sigma}+\nabla^\sigma\nabla_\beta
R_{\alpha\sigma}-\Box R_{\alpha\beta}-g_{\alpha\beta}\nabla_\sigma\nabla_\gamma
R^{\sigma\gamma}\nonumber\\
&-&2R_{\sigma\beta}
{R^\sigma}_\alpha+\frac{1}{2}g_{\alpha\beta}R_{\sigma\gamma}R^{\sigma\gamma}-\frac{1}{3}\Big(2\nabla_\alpha\nabla_\beta
R-2g^{\alpha\beta}\Box R\nonumber\\
&-&2RR_{\alpha\beta}+\frac{1}{2}g_{\alpha\beta}R^2\Big).
\end{eqnarray}
The Mannheim-Kazanas spherically symmetric solution to the Bach equation is given by the metric 
\begin{equation}
	{\rm d}s^{2}=-B(r)\, {\rm d}t^{2}+\frac{{\rm d}r^{2}}{B(r)}+r^{2}({\rm d}\theta^{2}+\sin^{2}\theta\,
	{\rm d}\phi^{2}) \label{metr}
\end{equation} 
in the usual Schwarzschild coordinates ($-\infty < t < \infty$, $r\geq0$, $0\leq\theta\leq\pi$ and $0\leq\phi\leq 2\pi$), where the lapse function $B(r)$ is defined as \cite{Mannheim:1989}
\begin{equation}\label{eq:originalWeylB(r)}
 B(r) = 1-\frac{\zeta (2-3\,\zeta\,\rho)}{r}  - 3\,\zeta\,\rho + \rho\, r - \sigma\, r^2. 
\end{equation}
The coefficients $\zeta$, $\rho$ and $\sigma$  are three-dimensional integration constants. The above solution reduces to the Schwarzschild-de Sitter solution for $\rho\rightarrow 0$ and therefore, at distances much smaller than $1/\rho$, it recovers general relativity. This solution has been also assessed for the Reissner--Nordstr\"{o}m spacetime in the presence of a charged source. In this context, the Weyl field equations become
\begin{equation}\label{eq:W=T}
   W_{\alpha\beta} = \frac{1}{4\mathcal{K}}~T_{\alpha\beta},
\end{equation}
in which $T_{\alpha\beta}$ is the energy-momentum tensor produced by the vector potential 
\begin{equation}\label{eq:vectorPotential}
    A_\alpha = \left(
    \frac{q}{r},0,0,0
    \right),
\end{equation}
with $q$ as the electric charge of the source \cite{Mannheim1991,Mannheim1991b}. In the same manner, in Ref.~\cite{Payandeh:2012mj}, {a reference lapse function of the form
\begin{equation}\label{eq:F_metric_0}
    B(r) = 1+\frac{1}{3}\left(c_2 r + c_1 r^2\right)
\end{equation}
was considered where the specification of the coefficients $c_1$ and $c_2$ was based on the weak field method. Accordingly, the last two terms of the above function can form a perturbation on the Minkowski spacetime, which can constitute the Poisson equation $\nabla^2 h_{\mu\nu} = 8\pi\mathcal{T}_{\mu\nu}$, with $h_{\mu\nu} = g_{\mu\nu}-\eta_{\mu\nu}$, that has the following 00 component:
\begin{equation}\label{eq:Poisson_1}
    \nabla^2 h_{00} = 8\pi\mathcal{T}_{00} = 8\pi\left( \frac{\tilde{m}}{\frac{4}{3}\pi \tilde{r}^3}+\frac{1}{8\pi}\frac{\tilde{q}}{r^4}\right),
\end{equation}
in which $\mathcal{T}_{00}$ is the scalar part of the energy-momentum tensor, corresponding to a charged spherically symmetric massive source of mass $\tilde{m}$, charge $\tilde{q}$ and radius $\tilde{r}$. Applying Eq.~\eqref{eq:Poisson_1} to the lapse function \eqref{eq:F_metric_0}, it is found \cite{Payandeh:2012mj}
{\begin{equation}\label{eq:c1c2}
    c_2 = -\frac{9\,r\,\tilde{m}}{\tilde{r}^3}-\frac{3}{2}\frac{\tilde{q}^2}{r^3}-3\,c_1 r,
\end{equation}}
substitution of which in Eq.~\eqref{eq:F_metric_0}, yields}
\begin{equation}
	B(r)=1-\frac{r^{2}}{\lambda^{2}}
	-\frac{Q^2}{4 r^2}, \label{lapse}
\end{equation}
in which
\begin{eqnarray}
&&\frac{1}{\lambda^2}=\frac{3\,\tilde{m}}{\tilde{r}^{3}}   +\frac{2\,c_1}{3},\label{par1}\\
&&Q=\sqrt{2}\,\tilde{q}.\label{par2}
\end{eqnarray}
For $\lambda>Q$, this spacetime allows for two horizons; the event horizon $r_+$ and the cosmological horizon $r_{++}$, given by (see appendix \ref{app:Ap})
\begin{eqnarray}
&& r_+= \lambda \sin\left( {1\over 2} \arcsin\left(\frac{Q}{\lambda} \right)  \right),\label{w.6}\\
&& r_{++}=\lambda \cos\left( {1\over 2} \arcsin\left(\frac{Q}{\lambda}\right)\right).\label{w.7}
\end{eqnarray}
The extremal black hole, characterized by the unique horizon $r_{\mathrm{ex}}=r_+=r_{++}=\lambda/\sqrt{2}$ is obtained for $\lambda=Q$. For $\lambda<Q$ the system encounters a naked singularity. Note that, letting $\tilde{r}$ to be the free radial distance, $3\tilde{m}\rightarrow 2 M$ and  ${2c_1}\rightarrow\pm\Lambda$ ($\Lambda$ is the cosmological constant), the lapse function in Eq.~\eqref{lapse} reduces to the Schwarzschild-(Anti-)de Sitter solution. Furthermore, the Reissner-Nordstr\"{o}m-(Anti-)de Sitter spacetime, is recovered by the imaginary transformation $Q\rightarrow 2 \,i\, Q_0$, in which $Q_0$ is the total charge of a spherical massive source. Accordingly, there is no trivial transition from the charged black hole proposed in Ref.~\cite{Payandeh:2012mj} to the general relativistic spherically symmetric spacetimes. 

We begin our study of the time-like geodesics in the next section, by constructing a Lagrangian formalism in the spacetime under study.

\section{The time-like geodesics around the charged Weyl black hole}\label{sec:TimeLikeStructure}

\begin{figure}[t]
	\begin{center}
		\includegraphics[width=9cm]{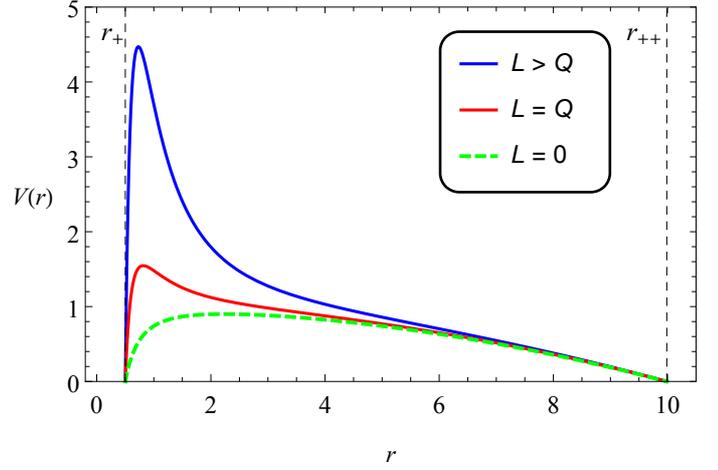}
	\end{center}
	\caption{The effective potential of the charged Weyl black hole, plotted for with $\lambda =10$ and $Q=1$, specified for particles with different designations of angular momentum. The larger the angular momentum, the more unstable is the potential's apex. The values of the horizons correspond to $r_+\simeq 0.5$ and $r_{++}\simeq 10$.}
	\label{fpot}
\end{figure}

The motion of massive particles in the spacetime given in Eq.~\eqref{metr} can be described by the Lagrangian \cite{Chandrasekhar:579245}
\begin{multline}\label{eq:lagrangian}
     2\mathcal{L} = \frac{1}{2}g_{\mu\nu} \dot x^\mu \dot x^\nu\\
     = \frac{1}{2}\left(
     -B(r) \dot t^2 + \frac{\dot r^2}{B(r)} + r^2\dot\theta^2 + r^2\sin^2\theta\dot\phi^2
     \right),
\end{multline}
in which, "dot" stands for differentiation with respect to the trajectory's affine parameter $\tau$. We can define the conjugate momenta 
\begin{equation}\label{eq:conjugate-pi}
    \Pi_\alpha = \frac{\partial\mathcal L}{\partial \dot x^\alpha},
\end{equation}
which according to the symmetries of the spacetime under consideration, leads to the two constants of motion  
\begin{equation}
	\Pi_{\phi}
	= r^{2} \dot{\phi} = L,\quad
	\textrm{and}\quad 
	\Pi_{t} = -B(r)\,\dot{t} = - E, 
	\label{w.11}
\end{equation}
where $L$ is the test particle's angular momentum (for unit of mass), and $E$ is an integration constant.  Here, $E$ cannot be regarded as the particles' energy because the spacetime is not asymptotically flat. Specifying the time-like geodesics by $2\mathcal{L} = -1$ and confining ourselves to motions on the equatorial plane ($\theta = \pi/2$), from Eqs.~\eqref{eq:lagrangian} and \eqref{w.11} we get
\begin{equation}\label{rtau}
    \dot r^2 = E^2-V(r),
\end{equation}
in which the gravitational effective potential of the system is defined as
\begin{equation}\label{poteff}
V(r)=B(r) \left( 1+\frac{L^2}{r^{2}}\right).
\end{equation}
The behavior of this potential for particles with different angular momentum has been plotted in Fig.~\ref{fpot}. As we can see, the intensity of the potentials' maximum is rather sensitive to $L$. The radial and angular motions of the test particles in this potential, are described by the following equations:
\begin{equation}\label{rt}
    \left(\frac{{\rm d}r}{{\rm d} t}\right)^{2}={ B^2(r)\over E^2}\left(E^2-V(r)\right),
\end{equation}
\begin{equation}\label{rphi}
    \left(\frac{{\rm d}r}{{\rm d}\phi}\right)^{2}= \frac{r^4}{L^2}\left( E^2-V(r)\right).
\end{equation}
The effective potential in Eq.~\eqref{poteff} is responsible for the determination of possible orbits around the black hole. 
In fact, the most essential feature of such potentials is their possibility of having any maximums or minimums. In the case of Fig.~\ref{fpot}, the potential expresses an instability at its maximum. The maximum apex in the potential, therefore, corresponds to unstable orbits or critical trajectories which will be discussed further in this paper. Other regions of the potential are as well, correspondents to different kinds of trajectories. 

In the next section, we will discuss the possible orbits in this potential, by presenting direct analytical solutions of the angular equations of motion.


\section{Angular Motion}\label{sec:angular}

\begin{figure}[t]
	\begin{center}
		\includegraphics[width=9cm]{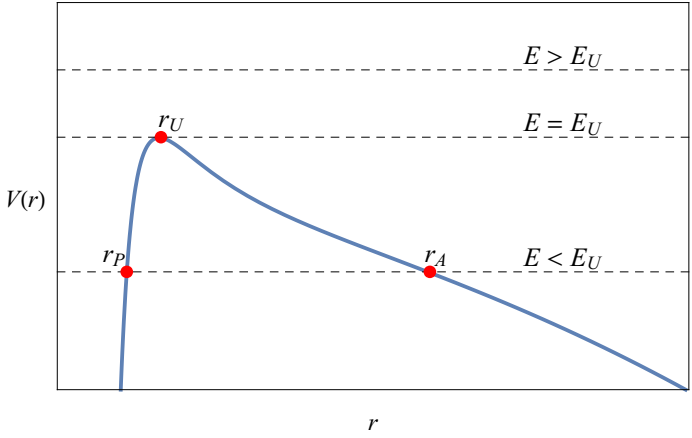}
	\end{center}
	\caption{The effective potential for test particles with angular momentum. Based on the values of $E$, several turning points (approaches) are available. These include the radius of unstable circular orbits $r_U$, and two other points, $r_P$ and  $r_A$. At these turning points, we have $E^2 = V(r_t)$.} 
	\label{fig:M_effectivePotential_0}
\end{figure}

In general, the most common trajectories followed by particles as they approach the black hole, are angular trajectories ($L\neq0$). Once again, we drag the reader's attention to the radial behavior of the effective potential, as illustrated in Fig.~\ref{fig:M_effectivePotential_0}. Corresponding to the values of $E$, the turning points $r_t$ relate to different kinds of orbits and they satisfy $E^2 = V(r_t)$. To determine these points, we should take care of their relevant orbital conditions. In fact, according to Fig.~\ref{fig:M_effectivePotential_0}, three turning points are denoted; $r_t = r_U$ (for unstable circular orbits), $ r_t = r_P$ (the smallest orbital separation) and $r_t = r_A$ (the largest orbital separation).
In the forthcoming subsections, we ramify the relevant orbital conditions of approaching test particles and determine the mentioned turning points in accordance with each particular type of motion. We begin with discussing the potential's maximum and its relevant quantities. Afterwards, other kinds of orbits are studied.


\subsection{Unstable circular orbits}\label{subsec:circular}

According to Fig.~\ref{fig:M_effectivePotential_0}, the effective potential offers instability in the motion of approaching particles, at points where $V'(r) = 0$ (prime stands for $\partial/\partial r$). Form Eq.~\eqref{poteff}, this generates
\begin{equation}\label{eq:Vprime=0}
    L^2 Q^2 - \left(2 L^2-\frac{Q^2}{2}\right)r^2-\frac{2}{\lambda^2}r^6 = 0,
\end{equation}
which is an equation of sixth order. Applying the Cardano's method, we can obtain three different radii for the unstable circular orbits, by solving Eq.~\eqref{eq:Vprime=0}. These read as (see appendix \ref{app:A})
\begin{eqnarray}
&&r_U=
\left(\Xi_0\sinh \left[\frac{1}{3}\arcsinh(\Xi_1)\right] \right)^{\frac{1}{2}}, \qquad L > \frac{Q}{2}\label{co1}\\
&&r_{U}=
\left( \frac{Q^4\lambda^2}{8}\right)^{\frac{1}{6}}, \qquad  \qquad \qquad \quad  \quad~~~~~~ L = \frac{Q}{2} \label{co2}\\
&&r_{U}=\left(\Xi_0 \cosh \left[\frac{1}{3}\arccosh(\Xi_1)\right] \right)^{\frac{1}{2}}, \qquad L < \frac{Q}{2} \label{co3}
\end{eqnarray}
where
\begin{eqnarray}
&&\Xi_0 = 4\lambda \sqrt{\frac{\left| L^2-Q^2/4\right| }{3}},\\
&&\Xi_1 = \frac{3 Q^2 L^2}{8\lambda} \sqrt{\frac{3}{\left| L^2-Q^2/4\right|^3}}.
\end{eqnarray}
One can also calculate the period of the above orbits, measured by the test particles (proper period) and a distant observer (coordinate period) \cite{Chandrasekhar:579245}. Exploiting Eqs.~\eqref{w.11}, we can obtain the following relations for a long-term circular orbit:
\begin{eqnarray}\label{eq:dt,dtau}
 \Delta\tau_U &=& \frac{r_U^2}{L_U}\,\Delta\phi_U,\label{eq:dtau}\\
    \Delta t_U &=& \frac{E_U}{L_U}\,\frac{r_U^2}{B(r_U)}\,\Delta\phi_U.\label{eq:dt}
\end{eqnarray}
For one complete orbit, we have $\Delta\phi_U = 2\pi$, and we define the proper and coordinate periods as
\begin{eqnarray}\label{eq:periods_definition}
     T_\tau &=& \frac{2\pi\, r_U^2}{L_U},\label{eq:periodProper}\\
    T_t &=& \frac{2\pi\, r_U^2 E_U}{B(r_U) L_U}.\label{eq:periodCoordinate}
\end{eqnarray}
The expression for $L_U$ is calculated by solving Eq.~\eqref{eq:Vprime=0} for the angular momentum at the fixed circular radius $r_U$. We have
\begin{equation}\label{eq:LU}
    L_U = \frac{1}{\sqrt{2}}\sqrt{\frac{4 r_U^4 - Q^2 \lambda^2}{\frac{Q^2 \lambda^2}{r_U^2}-2\lambda^2}}.
\end{equation}
This, together with the condition $E_U^2=V(r_U)$ at the distance $r_U$, provides
\begin{equation}\label{p2}
		T_\tau = 2\pi\lambda\,r_U\sqrt{\frac{4 r_U^2-2Q^2}{Q^2\lambda^2-4 r_U^4}}, 
\end{equation}
\begin{equation}\label{p1}
		T_{t} = \frac{4 \pi \lambda\,r_U^2}{\sqrt{\lambda^2 Q^2-4 r_U^4}}.
\end{equation}
Further in this section, we will discuss the critical trajectories corresponding to the above radii of unstable orbits. However for now, let us continue our discussion by studying the hyperbolic motions around the black hole.


\subsection{Orbits of the first kind and the scattering zone}

\begin{figure}[t]
		\begin{center}
			\includegraphics[width=7.5cm]{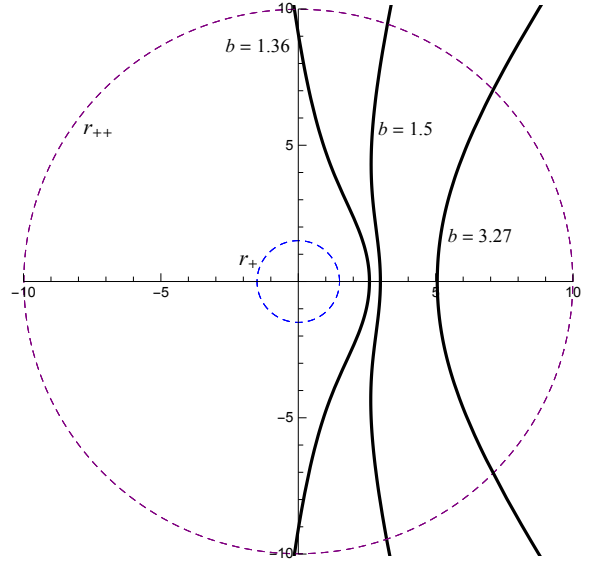}
		\end{center}
		\caption{Scattering of particles for different impact parameters $b=1.36, 1.5$ and $3.27$. It is observed that the scattering process can be attractive or repulsive, depending on the impact parameter. The plots have been done for $Q=1$ and $\lambda=10$. }
		\label{fig:scattering}
\end{figure}

In the case that, for orbiting test particles, the condition $E < E_U$ is satisfied, they can approach the black hole at two distinct points. Referring to Fig.~\ref{fig:M_effectivePotential_0}, these points are determined by $r_t = r_P$ and $r_t = r_A$, at which $\frac{\mathrm{d}r}{\mathrm{d}\phi}|_{r_t} = 0$ or $E^2 = V(r_t)$. The angular equation of motion in Eq.~\eqref{rphi} can be recast as
\begin{equation}\label{tl11}
\left(\frac{\mathrm{d}r}{\mathrm{d}\phi}\right)^2=\frac{r^6-\alpha \,r^4 - \beta\,r^2+\gamma}{L^2\lambda^2}\equiv\frac{\mathfrak{P}(r)}{L^2\lambda^2}, 
\end{equation}
where 
\begin{eqnarray}\label{eq:alpha_beta_gamma}
   &&\alpha=\lambda^2(1-E^2)-L^2,\\
   &&\beta=\lambda^2( L^2-Q^2),\\
   &&\gamma=\lambda^2 L^2Q^2.
\end{eqnarray}
the determination of the turning points $r_P$ and $r_A$ can be done by solving $\mathfrak{P}(r_t) = 0$ which is again an equation of sixth order and can be solved by means of the Cardano's method (see appendix \ref{app:B}). This results in
\begin{eqnarray}\label{eq:rP,rA}
&& r_A = 
		\left( \xi_0\cos \left[\frac{1}{3}\arccos(\xi_1)\right] +\frac{\alpha}{3}\right)  ^{1/2},\label{mr52a}\\
&& r_P =
		\left( \xi_0\cos \left[\frac{1}{3}\arccos(\xi_1)+\frac{4\pi}{3}\right] +\frac{\alpha}{3}\right)  ^{1/2},\label{mr52b}
\end{eqnarray}
where 
\begin{eqnarray}
    \xi_0 &=& 2\sqrt{\frac{\beta}{3}+\frac{\alpha^2}{9}},\\
    \xi_1&=&\left(\frac{8\alpha^3}{9}+4\alpha\beta-12\gamma\right)\sqrt{\frac{3}{(4\beta+\frac{4\alpha^2}{3})^3}}.
\end{eqnarray}
Particles reaching $r_A$ can experience a hyperbolic motion around the black hole and then escape to infinity. This kind of motion is known as orbit of the first kind (OFK) \cite{Chandrasekhar:579245} which has the significance of scattering. To find the explicit angular equation of motion for this process, we directly integrate Eq.~\eqref{tl11}, which results in (see appendix \ref{app:C})
\begin{equation}\label{mrd}
	r(\phi)=\frac{r_A}{\sqrt{4\wp(\varphi_A-\kappa_A\,\phi)+\frac{\beta r_A^2}{3\gamma}}},
\end{equation}
where $\wp(x)\equiv \wp(x; g_2, g_3)$ is the $\wp$-Weierstra$\ss$ function with
\begin{eqnarray}
g_2&=& \frac{r_A^4}{4}\left[\frac{\beta^2}{3\gamma^2}+\frac{\alpha }{\gamma}\right],\label{mr7ca}\\
	g_3&=& \frac{r_A^6}{16}\left[\frac{2\beta^3}{27\gamma^3}+\frac{\alpha \beta
	}{3\gamma^2}-\frac{1}{\gamma}\right],\label{mr7cb}
\end{eqnarray}
as its Weierstra$\ss$ coefficients. Additionally,
\begin{eqnarray}\label{eq:kappaP}
    \kappa_A&=&\frac{2Q}{r_A},\\
    \varphi_A&=&\ss\left(\frac{1}{4}-\frac{\beta r_A^2} {12\gamma}\right)
\end{eqnarray}
in which, $\ss(x)\equiv\wp^{-1}(x;g_2,g_3)$ is the inverse $\wp$-Weierstra$\ss$ function. The hyperbolic motion of particles around the black hole has been plotted in Fig.~\ref{fig:scattering}. Defining the impact parameter $b=L/E$, associated with the trajectories, we can see that the lower $b$ is, the more the trajectories are inclined to the black hole during their scattering.

\subsubsection{The scattering angle}

\begin{figure}[t]
		\begin{center}
			\includegraphics[width=8cm]{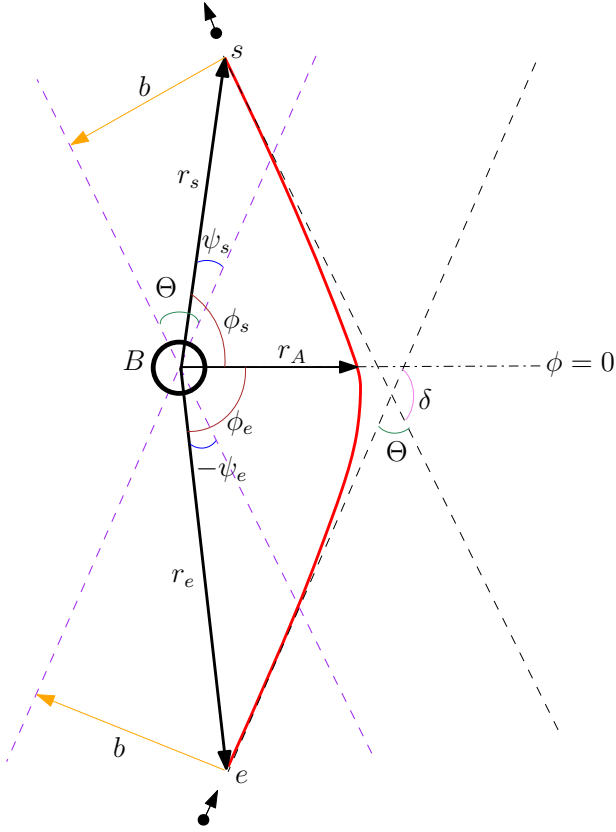}
		\end{center}
		\caption{A schematic illustration of the scattering phenomena. The shortest distance to the black hole $B$, has been taken to be $r_A$, lying on the $ \phi = 0$ line. The incident and the scattered particles are located respectively at $e(r_e,\phi_e,b)$ and $s(r_s,\phi_s,b)$.}
		\label{fig:scattering_Scheme}
\end{figure}

\begin{figure}[t]
		\begin{center}
			\includegraphics[width=8cm]{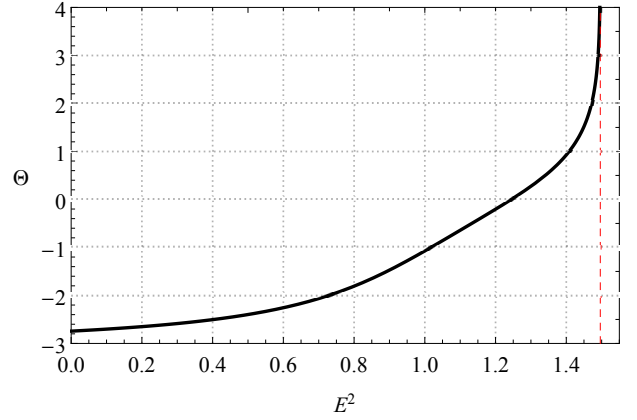}
		\end{center}
		\caption{The behavior of $\Theta$ in terms of $E^2$, demonstrated for $L=2$, $Q=1$ and $\lambda=10$. As it is expected, the scattering angle reaches its limit as $E$ tends to $E_U$ which in this case is around 1.496. }
		\label{fig:scatteringAngle}
\end{figure}

During the scattering process, the particles experience an escape to the infinity. Let us consider the scheme in Fig.~\ref{fig:scattering_Scheme}. The particles commence their approach to the black hole at point $e$ and the scattered particles recede to infinity at point $s$, which are characterized respectively by $e(r_e,\phi_e,b)$ and $s(r_s,\phi_s,b)$. Letting $r(\phi)|_{\phi=0} = r_A$, the shortest distance to the black hole is taken to be $r_A$, at which the scattering happens.
According to the figure, we have \cite{Fathi:2020}
\begin{equation}\label{eq:delta}
    \delta = \pi-\Theta = \phi_e - \psi_e + |\phi_s| - |\psi_s|.
\end{equation}
Any angle $\phi(r)$ observed by the moving particles in this kind of motion, is obtained by reversing Eq.~\eqref{mrd}, giving
\begin{equation}\label{eq:phi(r)}
    \phi(r) = \frac{1}{\kappa_A}\left[
    \ss\left(\frac{1}{4}-\frac{\beta\,r_A^2}{3\gamma}\right)
    -\ss\left(
    \frac{r_A^2}{4r^2}-\frac{\beta r_A^2}{12\gamma}
    \right)
    \right].
\end{equation}
Furthermore, according to the figure, it is easily inferred that
\begin{eqnarray}\label{eq:lensing2}
&&\psi_e = \Theta - \arcsin\left(\frac{b}{r_e}\right),\\
&&|\psi_s| = \Theta-\arcsin\left(\frac{b}{r_s}\right).
\end{eqnarray}
Assuming that the incident particles are coming from infinity and escaping to infinity, we have $\psi_e = |\psi_s| = \Theta$ and $\phi_e = |\phi_s|=\phi(\infty)\equiv\phi_\infty$. At this limit we can recast Eq.~\eqref{eq:delta} as $\Theta = 2\phi_\infty-\pi$, for which, applying Eq.~\eqref{eq:phi(r)}, we obtain the scattering angle as
\begin{equation}\label{mr.8}
\Theta=\frac{2}{\kappa_A} \left[  \ss\left({1\over 4}-{\beta \,r_A^2\over 12\gamma}\right)
-\ss\left(-{\beta \,r_A^2\over 12\gamma}\right)\right] -\pi.
\end{equation}
The evolution of the scattering angle has been plotted in Fig.~\ref{fig:scatteringAngle} which has an asymptotic behavior as $E\rightarrow E_U$.

\subsubsection{The differential cross section}

\begin{figure}[t]
		\begin{center}
			\includegraphics[width=8cm]{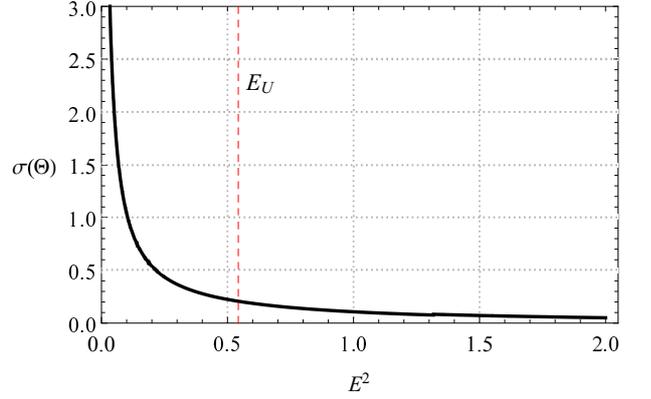}
		\end{center}
		\caption{The evolution of $\sigma(\Theta)$ in terms of $E^2$, plotted for $L=0.8$, $Q=0.5$ and $\lambda = 0.6$. For these values, $E^2_U \approx 0.54$. }
		\label{fig:differentialCrossSection}
\end{figure}

Regarding the spherical symmetry of our problem, the angle $\Theta$ obtained above, indeed measures the deflection angle between the incident and the scattered beams, that together with the azimuth angle $\phi$, can construct the solid angle element $\mathrm{d}\Omega = \sin\Theta\,\mathrm{d}\Theta\,\mathrm{d}\phi$ as the differential angular range of the scattered particles at angle $\Theta$. {Furthermore, since the impact parameter $b$ is perpendicular to the incoming and scattered trajectories, one can define the scattering cross section as the area covered by the scattered particles in the plane of $b$, which has the differential size $\mathrm{d}\sigma = b\,\mathrm{d}\phi \,\mathrm{d}b$. The differential cross section is then defined as 
\begin{equation}\label{eq:diffCross-def}
    \sigma(\Theta) \doteq \frac{\mathrm{d}\sigma}{\mathrm{d}\Omega} = \frac{b}{\sin\Theta} \left|\frac{\partial b}{\partial\Theta}\right|.
\end{equation}
} 
In fact, from Eq.~\eqref{mr.8} we have
\begin{equation}\label{eq:Theta_recast}
    \frac{\kappa_A}{2}(\Theta+\pi)=\varphi_{A_0} + \varphi_{A_1},
\end{equation}
in which
\begin{eqnarray}\label{eq:var1.var2}
    \varphi_{A_0} &\doteq& \ss\left(
    \frac{1}{4}-\frac{\beta\, r_A^2}{12\gamma}
    \right),\\ 
     \varphi_{A_1} &\doteq& -\ss\left(
   -\frac{\beta\, r_A^2}{12\gamma}
    \right).
\end{eqnarray}
We define
\begin{equation}\label{eq:Psi}
    \Psi(L) \doteq \wp\left(
     \frac{\kappa_A}{2}(\Theta+\pi)
    \right) = \wp\left(
    \varphi_{A_0} + \varphi_{A_1}
    \right),
\end{equation}
where \cite{handbookElliptic}
\begin{equation}\label{eq:Psi(L)}
    \Psi(L) = \frac{1}{4}\left[
    \frac{\wp'(\varphi_{A_0})-\wp'(\varphi_{A_1})}{\wp(\varphi_{A_0})-\wp(\varphi_{A_1})}
    \right]^2-\wp(\varphi_{A_0})-\wp(\varphi_{A_1}),
\end{equation}
in which the differentiation of the Weierstra$\ss$ function is defined as
\begin{equation}\label{eq:diff-wp-def}
    \wp'(x)\equiv\frac{\mathrm{d}}{\mathrm{d}x}\wp(x) = - \sqrt{4\wp^3(x)-g_2\wp(x)-g_3}.
\end{equation}
Note that, using the definition in Eq.~\eqref{eq:Psi(L)}, we can recast Eq.~\eqref{eq:diffCross-def} as
\begin{multline}\label{eq:diffCross-def_1}
    \sigma(\Theta) = b\csc\Theta\left|
    \frac{\partial\Psi}{\partial\Theta}
    \right|
    \left|
    \frac{\partial b}{\partial\Psi}
    \right|\\
    = \frac{\kappa_A}{4}\csc\Theta\left|
    \wp'\left(
    \frac{\kappa_A}{2}(\theta+\pi)
    \right)
    \right|
    \left|
    \frac{\partial b^2}{\partial\Psi}
    \right|,
\end{multline}
for which, considering $\frac{\partial b^2}{\partial\Psi} = \frac{\partial b^2/\partial L}{\partial\Psi/\partial L}$, we finally obtain 
\begin{equation}\label{eq:sigmaTheta_2}
    \sigma(\Theta) = \frac{\kappa_A L}{2E^2}\csc\Theta\left|
    \wp'\left(
    \frac{\kappa_A}{2}(\theta+\pi)
    \right)
    \right|
    \left|
    \frac{\partial\Psi}{\partial L}
    \right|^{-1}.
\end{equation}
The complexity of the relation of $\Psi(L)$, makes the resultant expression of $\sigma(\Theta)$ rather large and complicated. We however, have demonstrated the behavior of this function in Fig.~\ref{fig:differentialCrossSection}, in terms of the quantity $E^2$. We have considered smaller values for the constants to be able to generate a more perceptible plot. Note that, there is an asymptotic behavior as $E\rightarrow0$, and $\sigma(\Theta)$ tends to zero, soon after $E$ passes $E_U$.

\subsubsection{Radial acceleration} 

\begin{figure}[t]
	\begin{center}
		\includegraphics[width=80mm]{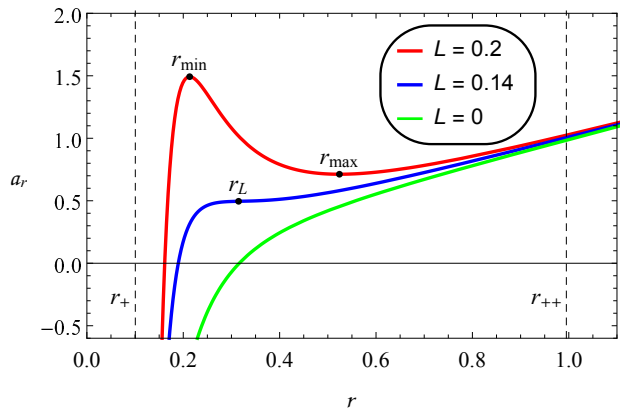}
	\end{center}
	\caption{The evolution of the radial acceleration $a_r\equiv\ddot{r}$ inside the casual region $r_+<r<r_{++}$ plotted for $Q=0.2$, $\lambda=1$ and three different values of $L$. The case of $L=0.2$ has two extremums at $r_{\mathrm{min}} \approx 0.21$ and $r_{\mathrm{max}} \approx 0.52$. The case of $r_{\mathrm{min}} = r_{\mathrm{max}} = r_L$ happens for $L=0.14$ where $r_L \approx 0.31$.}
	\label{fig2}
\end{figure}

The equation of motion for the radial coordinate in Eq.~\eqref{rtau}, beside demonstrating the way through which the particles approach the black hole, can also provide information on the Newtonian centripetal effective force acting on the particles. This force is indeed indicated by the radial acceleration $a_r$ which is defined as $a_r\equiv \ddot r$ in terms of the radial coordinate. Using Eq.~\eqref{rtau} we have
\begin{equation}\label{1.18}
a_r=-\frac{1}{2}V'(r) = -\frac{L^2 Q^2}{2 r^5}+\frac{L^2-Q^2/4}{r^3}+\frac{r}{\lambda^2}.
\end{equation} 
Introducing $r_{\mathrm{max}}$ and $r_{\mathrm{min}}$, respectively as the turning points where $a_r$ reaches its maximum and minimum (by satisfying $\partial_r a_r=0$), we obtain
\begin{eqnarray}
&& r_{\mathrm{max}}=
\left( \eta_0\cos \left[\frac{1}{3}\arccos(\eta_1)\right]\right)  ^{1/2},\label{rM}\\
&& r_{\mathrm{min}}=
\left( \eta_0\cos \left[\frac{1}{3}\arccos(\eta_1)+\frac{4\pi}{3}\right] \right)  ^{1/2},\label{rm}
\end{eqnarray}
where 
\begin{eqnarray}\label{eq:eta0_1}
 &&\eta_0=2\lambda\sqrt{L^2-\frac{Q^2}{4}},\\
 &&\eta_1=-\frac{5 L^2 Q^2}{4\lambda} \left(L^2-\frac{Q^2}{4}\right)^{-\frac{3}{2}},
\end{eqnarray}
and are valid only for $Q<2L$. These distances have the identical value $r_L$ (corresponding to $\eta_1 = \pm 1$), when the angular momentum approaches the value $L_0$ given by 
\begin{equation}\label{L0}
L_0= \sqrt{
\chi_1+ \chi_2 \cosh \left[\frac{1}{3}\arccosh\left[ \frac{\chi_3}{\chi_2^3}\right] \right]},
\end{equation}
where
\begin{eqnarray}
&& \chi_1=\frac{9 Q^2}{4},  \\
&& \chi_2=\frac{20 Q}{8\sqrt{3}\,\lambda},  \\
&&  \chi_3=\frac{25 Q^4}{1024 \lambda^2}.   
\end{eqnarray}
The equality $r_{\mathrm{max}}\equiv r_{\mathrm{min}} = r_L$ has been shown in Fig.~\ref{fig2}, where we  have plotted $a_r$ for three different values of $L$. In accordance with the values chosen in the figure, the $L=0.14$ curve has only one extremum corresponding to $r_L\approx 0.31$. In this case, the test particles will experience a constant effective force towards the black hole while traveling on their trajectories.

In this subsection, we scrutinized the properties of the hyperbolic trajectories followed by scattered test particles. However, altering the point of approach from $r_A$, particles of the same impact parameter may experience a different fate. This is what we will study in the next subsection.

\subsection{Orbits of the second kind}

\begin{figure}[t]
	\begin{center}
		\includegraphics[width=7cm]{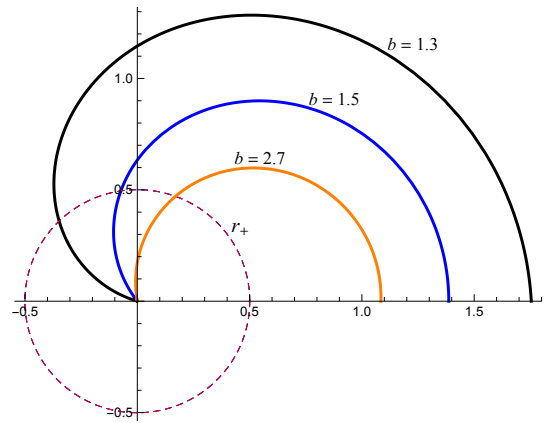}
	\end{center}
	\caption{Orbits of the second kind for particles approaching the black hole at $r=r_P$, for three different impact parameters, $b = 1.3, 1.5$ and $2.7$. As we can see, smaller impact parameters in this kind of orbit result in larger paths for the orbiting particles before their fall into the event horizon, and therefore, a more intense change in the shape of orbit in the final segment. The plots have been done for $Q=1$ and $\lambda=10$.}
	\label{fig4}
\end{figure}

The deflecting trajectories corresponding to the case of $E<E_U$, can also occur once the approaching point to the black hole coincides with the turning point $r_P$ in Eq.~\eqref{mr52b} ($r_+<r_P<r_U$). From this point, the test particles are dragged into the event horizon and therefore follow an orbit of the second kind (OSK) \cite{Chandrasekhar:579245}. Pursuing the same method, applied in deriving the equation of motion for the OFK, we obtain
\begin{equation}\label{mra}
r(\phi)=\frac{r_P}{\sqrt{4\wp(\varphi_P+\kappa_P\phi)+\frac{\beta r_P^2}{3\gamma}}},
\end{equation}
with the corresponding Weierstra$\ss$ coefficients
\begin{eqnarray}
&&g_{22}= \frac{r_P^4}{4}\left[\frac{\beta^2}{3\gamma^2}+\frac{\alpha }{\gamma}\right],\label{mr8ca}\\
&&g_{33}= \frac{r_P^6}{16}\left[\frac{2\beta^3}{27\gamma^3}+\frac{\alpha \beta
 }{3\gamma^2}-\frac{1}{\gamma}\right],\label{mr8cb}
\end{eqnarray}
and
\begin{eqnarray}\label{eq:kappaA}
&&\kappa_P=\frac{2Q}{r_P},\\
&& \varphi_P=\ss\left(\frac{1}{4}-\frac{\beta r_P^2} {12\gamma}\right).
\end{eqnarray}
In Fig.~\ref{fig4} we have demonstrated the OSK for particles with three different impact parameters. The larger the impact parameter is, the more the trajectories need to curve in their final segment, before their in-fall to the black hole. 

Now that the deflecting trajectories have been discussed, in the next section, we pay attention to the case that the particles' impact parameter raise to that of unstable circular orbits. This kind of orbit, corresponds to the critical trajectories.

\subsection{Critical trajectories }

\begin{figure}[t]
	\begin{center}
		\includegraphics[width=8.5cm]{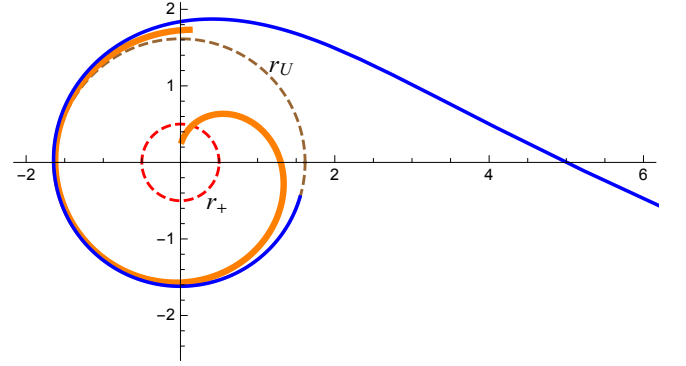}
	\end{center}
	\caption{The critical trajectories $r_I(\phi)$ (blue) and $r_{II}(\phi)$ (orange) plotted for $Q=1$, $\lambda=10$ and $L=2$. For this values, $E_U \approx 1.5 $ and $r_U \approx 1.6$ and the trajectories have been plotted for $\tilde{R} \approx 7.67$ and $\tilde{R}_0 = 1.3$. }
	\label{fig:critical}
\end{figure}

In the case of $E=E_U$, the particles can be confined on unstable circular orbits of the radius $r_U$. This kind of motion is indeed ramified into two cases; critical trajectories of the first kind (CFK) in which the particles come from a distant position $\tilde{R}$ to $r_U$ and those of the second kind (CSK) where the particles start from an initial point $\tilde{R}_0$ at the vicinity of $r_U$ and then tend to this radius by spiraling. Applying the angular equation of motion and pursuing the same methods as in the case of deflecting trajectories, we obtain the following equations of motion for the aforementioned trajectories:
\begin{equation}\label{crt1a}
    r_I(\phi)=\frac{\tilde{R}}{ \sqrt{(1+\frac{\tilde{R}^2}{r_U^2}) \tanh^2\left(\varphi_{C_1}+\kappa_{C} \phi\right)-1}}
\end{equation}
for the CFK, and
\begin{equation}\label{crt1b}
    r_{II}(\phi)=\frac{\tilde{R}_0}{ \sqrt{(1+\frac{\tilde{R}_0^2}{r_U^2}) \tanh^2\left(\varphi_{C_2}+\kappa_{C} \phi\right)-1}}
\end{equation}
for the CSK. Here, 
\begin{eqnarray}\label{eq:coefficnents_critical}
&&\kappa_{C} = \frac{r_U \sqrt{\tilde{R}^2+r_U^2}}{\lambda L},\\
&&\varphi_{C_1}=\mathrm{arctanh}\left(\frac{r_U}{\sqrt{\tilde{R}^2+r_U^2}}\right),\\
&&\varphi_{C_2} = \mathrm{arctanh}\left(
\frac{r_U \sqrt{\tilde{R}^2+\tilde{R}_0^2}}{\tilde{R}_0 \sqrt{\tilde{R}^2+r_U^2}}
\right).
\end{eqnarray}
In Fig.~\ref{fig:critical}, the CFK and CSK have been demonstrated in a single figure to indicate their difference in approach to the region of the circular orbits. 

Note that, if the parameter $E$ of the particles is raised to values larger than $E_U$, the trajectories can no more maintain any kinds of stability and they fall into the event horizon. This kind of motion is discussed in the next subsection. 

\subsection{Capture zone}

\begin{figure}[t]
	\begin{center}
		\includegraphics[width=7.5cm]{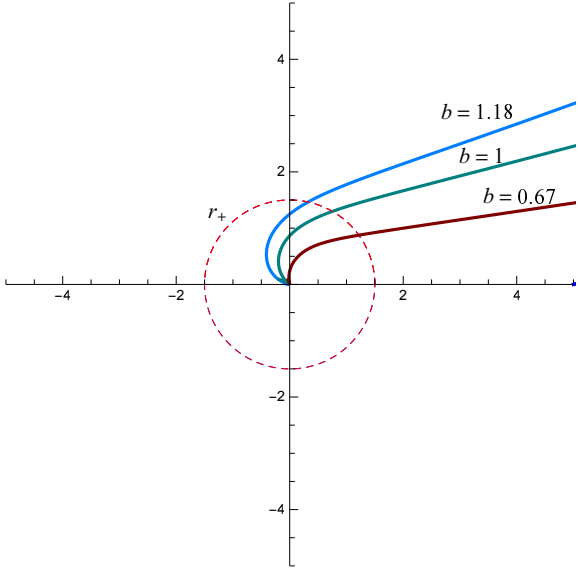}
	\end{center}
	\caption{The captured trajectories for particles approaching from $r_0=5$, plotted for $Q=1$, $\lambda=10$ and $L=2$. Accordingly, the critical impact parameter is $b_U \approx 2 $ and the trajectories plotted here correspond to $b = 1.18, 1$ and $0.67$.}
	\label{fig:capture}
\end{figure}

In addition to the OSK, terminating orbits can also occur when the value of $E$ for the approaching particles exceeds that of unstable circular orbits; i.e. $E>E_U$. If we consider approaching particles with the same angular momentum, this corresponds to particles with $b<b_U$, where $b_U=L_U/E_U$ is the critical impact parameter possessed by particles traveling on the unstable circular orbits. The equation of captured trajectories is similar to that for the deflecting trajectories and is obtained by replacing $r_A$ or $r_P$ by a constant initial distance, say $r_0$, as an arbitrary starting point. This kind of motion, has been plotted in Fig.~\ref{fig:capture} for three different impact parameters in the allowed range.\\

In this section, we studied the possible types of angular motion for particles with different impact parameters and calculated analytically, the equations of motion for the corresponding trajectories. We showed that the particles can escape the black hole region and although the effective potential does not allow for planetary orbits, nevertheless, the test particles can be confined in circular orbits outside the event horizon. In all of these cases, the angular momentum plays a crucial role, without which, any approaching particle will inevitably fall into the black hole. Although this kind of motion does not absorb the interest regarding the types of orbit (because no orbits are available), however, there are some interesting relativistic effects according to the concept of time which are worth discussing. These materials are dealt with in the next section.

\section{Radial Trajectories}\label{sec:radial}

The study of radial trajectories of falling particles, beside its historical root in the  Newtonian description of gravity, has numerous advantages in investigating the world-line structure of black hole spacetimes. For example, one can discuss the gravitational clock effect for falling observers in gravitating regions, which is also tightly related to the gravitational redshift of light rays passing black holes. Another interesting subject to discuss, is the phenomenon of a \textit{frozen star} \cite{Zeldovich:2014} which is related to the differences in the time measurements, done by distant observers and falling ones (for text book reviews, see for example Ref.~\cite{Ryder:2009}). In this section, a similar phenomenon will be studied for radially moving particles in the exterior geometry of a charged Weyl black hole. 

The radial motion of particles is characterized by the condition $L=0$, for which the effective potential reduces to
\begin{equation}\label{rt1} 
V_r(r)= 1-\frac{r^2}{\lambda ^2}-\frac{Q^2}{4 r^2}.
\end{equation}
which allows a maximum at $r_u = \sqrt{Q\lambda/2}$, having the value
\begin{equation}\label{eq:Eu2}
   V_r(r_u) \equiv E_u^2 = 1-\frac{Q}{\lambda}.
\end{equation}
Before going any further, let us ramify the types of possible radial motions, based on the value of $E^2$ compared with the above $E^2_u$. \begin{itemize}
	\item \textit{Frontal  scattering}: When $E<E_u$, particles approaching the black hole from a finite distance, are diverted at $r_a$ (or $r_p$) towards the black hole's horizons. Since no angular motion is considered for the particles, this kind of scattering is completely frontal. 
	
	\item \textit{Critical radial motion}: For $E=E_u$, particles can stay on an unstable radial distance of radius $r=r_u$.  Therefore, particles coming from an initial distance $r_i$ or $d_i$ ($r_u < r_i < r_{++}$ and $r_+ < d_i < r_u$) will ultimately fall on $r_u$.
	
		\item \textit{Radial capture}: If $E>E_u$, particles coming from a finite distance $\rho_0$ ($r_+ < \rho_0 < r_{++}$), are pulled towards the horizons from the same distance. 
\end{itemize} 
Further in this section, we will study these types of radial trajectories which are classified in terms of $E$. For now, let us rewrite the radial velocity relations given in Eqs.~\eqref{rtau} and \eqref{rt} as
\begin{eqnarray}
&&\left(\frac{{\rm d}r}{{\rm d}\tau}\right)^{2}=\frac{
	r^4+(E^2-1) \lambda^2 r^2+\frac{Q^2\lambda^2}{4}}{\lambda^2r^2}
\equiv \frac{\mathfrak{p}(r)}{r^2}, \label{vel1}\\
&&\left(\frac{{\rm d}r}{{\rm d}t}\right)^{2}=\frac{
(r^2-r^2_{+})^2(r^2_{++}-r^2)^2\,\mathfrak{p}(r)}{E^2 \lambda^4 r^6}.\label{vel2}
\end{eqnarray}
These are the key relations in scrutinizing the radial trajectories of different kinds. In this section, the possible motions are studied regarding the time measurements done by observers comoving with the trajectories ($\tau$) and distant observers ($t$).

\subsection{Frontal  scattering}

\begin{figure}[t]
	\begin{center}
		\includegraphics[width=8cm]{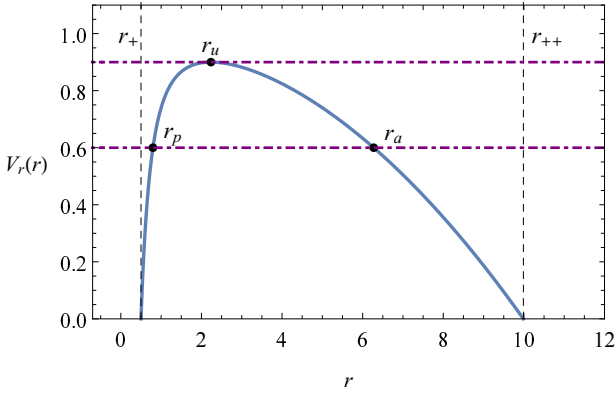}
	\end{center}
	\caption{The effective potential for radial trajectories plotted for $Q=1$ and $\lambda=10$. The radial distances $r_u$, $r_p$ and $r_a$ have been indicated.}
	\label{fig1}
\end{figure}

\begin{figure}[t]
	\begin{center}
		\includegraphics[width=8cm]{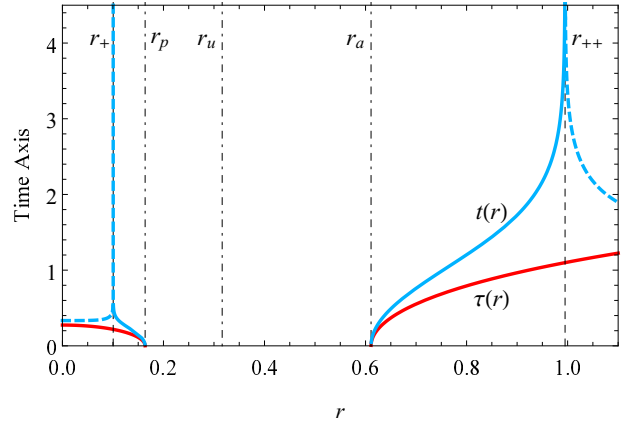}
	\end{center}
	\caption{The radial behavior of the proper and coordinate times in the two kinds of frontal scattering, plotted for $Q=0.2$, $\lambda = 1$ and $E^2 = 0.6$. After being scattered from $r_a$ (or $r_p$), the comoving observers see a horizon crossing. This is while a distant observer never observe this (frozen falling particles).}
	\label{fig:radialScatter}
\end{figure}

As we discussed in the previous section, the black hole
allows for scattering of angular geodesics. This also holds for radial trajectories when the condition $E<E_u$ is satisfied. Similarly, two turning points are available at either sides of $r_u$, namely $r_p < r_u <r_a$ (see Fig.~\ref{fig1}). Since they are turning points, these distances are identified by solving $\mathfrak{p}(r) = 0$, from which we obtain
\begin{equation}\label{rp}
r_p=  
\lambda \sqrt{1-E^2}\sin\left( {1\over 2} \arcsin\left( \frac{1-E_u^2}{1-E^2}
\right) \right),
\end{equation}
\begin{equation}\label{ra}
r_a= 
\sqrt{1-E^2}\cos\left( {1\over 2} \arcsin\left( 
\frac{1-E_u^2}{1-E^2}
\right)\right).  
\end{equation}
In the case of $E=0$, the above radial distances tend to the event and cosmological horizons. In  Fig.~\ref{fig1}, the effective potential $V_r(r)$ has been plotted, where the extremum $r_u$ and the turning points $r_p$ and $r_a$ are indicated.
Since these turning points are solutions to $\mathfrak{p}(r) = 0$, we can therefore rewrite Eq.~\eqref{vel1} as
\begin{equation}\label{velpropruth}
\left(\frac{{\rm d}r}{{\rm d}\tau}\right)^{2}=\frac{(r^2-r_a^2)(r^2-r_p^2)}{r^2}
\equiv \frac{\mathfrak{p}_{s}(r)}{\lambda^2 r^2},
\end{equation}
which implies $\mathfrak{p}(r)=\mathfrak{p}_s(r)/\lambda^2$. The first kind of scattering, happens when the particles approach at $r_a$. Let us assume that for comoving and distant observers, the particles are at $r=r_a$, when $\tau=t=0$. Accordingly, exploiting Eqs.~\eqref{velpropruth} and \eqref{vel2}, we obtain the following radial dependencies for the time parameters:
\begin{equation}\label{tauradsct1}
\tau(r)={\lambda\over 2}\ln \left| {2\left(\sqrt{\mathfrak{p}_s(r)}
	+ r^2\right)-(1-E^2)\over 2 r_a^2-(1-E^2)}\right| 
\end{equation}
for the comoving, and
\begin{equation}
t(r)={\lambda^3 E \over 2(r^2_{++}-r^2_+)} \left[{r^2_{++}\ln \left| F_1(r)\right|\over \sqrt{\mathfrak{p}_s(r_{++})}}-{r^2_{+}\ln \left| F_2(r)\right|\over \sqrt{\mathfrak{p}_s(r_+)}} \right]
\label{teradsct}
\end{equation}
for the distant observers, where
\begin{eqnarray}
F_1(r)&=&  {(r^2_{++}-r^2_a)\over (r^2_{++}-r^2)}
{F_{++}(r)\over F_{++}(r_a)},
\\
F_2(r)&=& {(r^2_a-r^2_{+})\over (r^2-r^2_{+})}
{F_{+}(r)\over F_{+}(r_a)},
\end{eqnarray}
in which, 
\begin{eqnarray}
F_{++}(r) &=&  2\mathfrak{p}_s(r_{++})+
(1-E^2-2r^2_{++}) (r^2_{++}-r^2)
\nonumber\\
&&+2\sqrt{\mathfrak{p}_s(r_{++})\,P_{++}(r)},\\
F_{+}(r) &=&  2\mathfrak{p}_s(r_{+})-
(1-E^2-2r^2_{+}) (r^2-r^2_{+})
\nonumber\\
&&+2\sqrt{\mathfrak{p}_s(r_{+})\,P_{+}(r)},
\end{eqnarray}
and
\begin{eqnarray}
P_{++}(r) &=& \mathfrak{p}_s(r_{++})+
(1-E^2-2r^2_{++}) (r^2_{++}-r^2)
\nonumber\\
&&+(r^2_{++}-r^2)^2,\\
P_{+}(r) &=& \mathfrak{p}_s(r_{+})-
(1-E^2-2r^2_{+}) (r^2-r^2_{+})
\nonumber\\
&&+(r^2-r^2_{+})^2.
\end{eqnarray}
To obtain the radial behavior of the time parameters in the second kind scattering (scattering from $r_p$), it suffices to exchange $r_a\rightarrow r_p$ in the above relations and reverse the evolution. In Fig.~\ref{fig:radialScatter}, the radial behaviors of $t(r)$ and $\tau(r)$ have been plotted for a specific value of $E$ for the two kinds of scattering. As we can see, the comoving observers see particles crossing the horizons, whereas, according to the distant observers, the particles will never cross the horizons. In this regard, at the vicinity of the horizons, the particles appear \textit{frozen} to the distant observers. 

In the next subsection, we consider that particles travel in the effective potential's maximum.

\subsection{Critical radial motion}

\begin{figure}[t]
	\begin{center}
		\includegraphics[width=80mm]{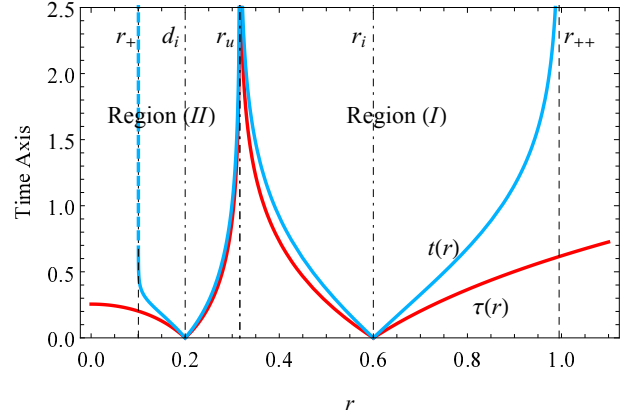}
	\end{center}
	\caption{Plot of the  critical radial motion in regions ($I$) and ($II$), plotted for $Q=0.2$, $\lambda = 1$ and $E^2=0.8$. It is assumed $r_i = 0.6$ and $d_i = 0.2$. In both cases, the comoving and distant observers see that the particles approach $r_u$ asymptotically, whereas once again, the horizon crossing is seen only for comoving observers.}
	\label{fig:radialCritical}
\end{figure}

Motion of particles with $E = E_u$, coming from $r_i>r_u$
or $d_i<r_u$ (respectively, regions ($I$) and ($II$) in Fig.~\ref{fig:radialCritical}), depends on the initial conditions at these points. According to Fig.~\ref{fig:radialCritical}, the discontinuity of $\frac{\mathrm{d}\tau}{\mathrm{d}r}$ and $\frac{\mathrm{d}t}{\mathrm{d}r}$, at $r_i$ and $d_i$, tell us about the final fate of the approaching particles. In this regard, they can either fall on $r = r_u$ or be pulled towards the horizons. Both fates can be obtained by integrating the equations of motion for the time parameters. For particles coming from $r_i$, we derive the following temporal relations in accordance with the comoving and distant observers:
\begin{eqnarray}
&&\tau_I (r)=\pm {\lambda\over 2} \ln\left| {r^2-r_u^2\over r_i^2-r_u^2}
\right|, \label{mrc1}\\
&&t_I (r)=\pm {\lambda^3 E\over 2} \left[ t_u (r)-t_{++} (r)-t_+(r)\right],\label{mrc2}
\end{eqnarray}
where
\begin{eqnarray}
&&t_{++}(r)={r^2_{++}\over (r^2_{++}-r^2_+)(r^2_{++}-r^2_u)} \ln\left| {r^2_{++}-r^2\over r_{++}^2-r_i^2}
\right|,\\
&&t_{+}(r)={r^2_{+}\over (r^2_{++}-r^2_+)(r^2_{u}-r^2_+)} \ln\left| {r^2-r_+^2\over r_i^2-r_+^2}
\right|,\\
&&t_{u}(r)={r^2_{u}\over (r^2_{++}-r^2_u)(r^2_{u}-r^2_+)} \ln\left| {r^2-r_u^2\over r_i^2-r_u^2}
\right|.
\label{mrc2a}
\end{eqnarray}
The corresponding evolution of these coordinates has been demonstrated in Region ($I$) of Fig.~\ref{fig:radialCritical}. The temporal equations of motion for particles coming from $d_i$ are  similar to the last ones and are given by considering the exchanges $\tau_{II}(r)=-\tau_I(r)$, $t_{II}(r)=-t_I(r)$ and $r_i\rightarrow d_i$. Region ($II$) of Fig.~\ref{fig:radialCritical}, indicates their radial evolution.

The cases stated here, constitute the characteristics of the critical radial motions around the black hole when the particles are subjected to the maximum of the radial effective potential. As it is noticed, when the initial conditions are satisfied, comoving observers see a horizon crossing whereas for distant observes the particles will never cross the horizons. In the next section, we consider the case in which the particles travel in a potential which exceeds the mentioned maximum. 

\subsection{Radial capture}

\begin{figure}[t]
	\begin{center}
		\includegraphics[width=8cm]{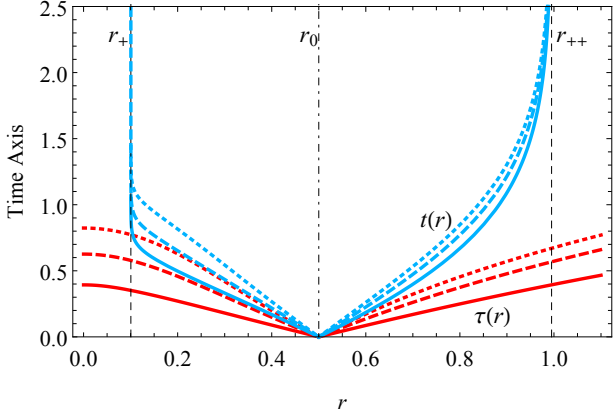}
	\end{center}
	\caption{Plot of the radial capture for particles. With  $Q=0.2$, $\lambda=1$  and $\rho_0=0.5$. The way of the behavior of the time parameters are similar to those in the radial scattering. The plots have been done for three different values of $E>E_u$ and are classified as \textbf{dotted}: $E^2 = 1 < 1+Q/\lambda$, \textbf{dashed}: $E^2=1.2 = 1+Q/\lambda$ and \textbf{solid}: $E^2=2 > 1+Q/\lambda$.}
	\label{fig:radialCapture}
\end{figure}

In the case that $E>E_u$, the particle trajectories are inevitably pulled towards the horizons; the particles are captured. To solve Eq.~\eqref{vel1} for the comoving time parameter, we consider a reference value $E=1+Q/\lambda$ which is in general, larger than $E_u$. If we assume that at $\tau = 0$, the particles are at a finite distance $\rho_0$ (i.e. $\tau(\rho_0) = 0$), then the solutions are classified as
\begin{itemize}

\item For $E_u^2<E^2<1+\frac{Q}{\lambda}$:
\begin{equation}
    \tau(r)=\pm{ \lambda\over 2} \left[  \arcsinh \left( {2 r^2 +E^2-1\over \eta_E}\right) -k_0\right].
\end{equation}
\item For $E^2 = 1+\frac{Q}{\lambda}$:
\begin{equation}
    \tau(r)=\pm{ \lambda\over 2} \ln \left|  { 2r^2+Q\over 2\rho_0^2+Q}  \right|.
\end{equation}
\item For $E^2>1+\frac{Q}{\lambda}$:
\begin{equation}
   \tau(r)=\pm{ \lambda\over 2} \ln \left|  { 2\sqrt{\mathfrak{p}(r)}+2r^2+E^2-1\over 2\sqrt{\mathfrak{p}(r)}+2\rho_0^2+E^2-1}  \right|.
\end{equation}

\end{itemize}
In above, we have defined
\begin{eqnarray}
  &&\eta_E=\sqrt{(E^2-E_u^2)(1+\frac{Q}{\lambda}-E^2)},\\
  &&k_0=\arcsinh \left({2 \rho_0^2 +E^2-1\over \eta_E}\right).
\end{eqnarray}
The relation of the time parameter for the distant observers can be considered the same as that in Eq.~\eqref{teradsct}, and we just need to replace $r_a\rightarrow \rho_0$. In Fig.~\ref{fig:radialCapture} we have plotted the behavior of the above coordinates in the radial capture process. The behavior is more or less like the radial scattering, except the fact that in both kinds of trajectories (towards $r_{++}$ or $r_+$), the trajectories are being captured from the initial distance $\rho_0$.\\

In this section, we presented a detailed study of the radial trajectories and the evolution of the time parameters, and the concept of horizon crossing was demonstrated by analyzing different types of motion. So far, the world-line structure of moving particles around the black hole has been investigated by calculating the equations of motion in connection with specific initial conditions. To continue with our discussion and as the last subject, we discuss a different, yet quite interesting impact of spacetime curvature around massive objects. In this regard, in the next section, we study a classical test, according to which, the spacetime effect on the spin vector of an orbiting gyroscope is discussed.

\section{Geodetic precession}\label{sec:precession}

In 1916, de Sitter imposed a relativistic correction to the gyroscopic precession of the Earth-moon system in its orbiting motion in the curved spacetime around the sun \cite{deSitter:1916}. This correction, known as geodetic effect (or geodetic precession, de Sitter precession or de Sitter effect), does not take into account the rotation of the central mass. The inclusion of this latter for rotating objects, results in a more general effect, called the dragging of inertial frames (or the Lense-Thirring effect) \cite{Lense:1918}. The geodetic precession effect has had a great influence in astrophysical observations and in fact constitutes one of the significant tests of general relativity. From a theoretical viewpoint, however, there are several methods in the derivation of geodetic precession and frame dragging (for alternative derivations and reviews see for example Refs.~\cite{Schiff:1960,Ashby:1990,Krisher:1997,Jonsson:2007,Wohlfarth:2013,lammerzahl:2001,Will:2014}). Here, we pursue a well-known method, consisting of a transformation to the local frame of an orbiting gyroscope in the curved spacetime generated by metric potential \eqref{lapse}. Same method has been employed in Ref. \cite{Said:2013} to calculate the geodetic precession in the Mannheim-Kazanas solution of the conformal Weyl gravity. Other methods, including the parameterized post-Newtonian (PPN) formalism can be found extensively in the available literature (see for example Ref.~\cite{Misner:1973}).  

Now we calculate the geodetic precession of the spin vector $\bar{\bm{S}}$ of a gyroscope angular which is orbiting with the angular velocity $\omega$. To proceed with this, let us identify the local frame of the gyroscope, by introducing the rotating coordinate system, characterized by the new angular coordinate
\begin{equation}\label{eq:varphi}
    \mathrm{d}\varphi = \mathrm{d}\phi - \omega\, \mathrm{d}t.
\end{equation}
This changes the non-rotating metric \eqref{metr} to that in rotating coordinates, which for $\theta = \pi/2$ reads as
\begin{multline}\label{eq:rotatingFrame}
    \mathrm{d}s^2 =-\left( B(r)-r^2\omega^2\right) \left( \mathrm{d}t-\frac{r^2\omega}{B(r)-r^2\omega^2}\,\mathrm{d}\varphi\right)^2 \\
 +\frac{\mathrm{d}r^2}{B(r)}
+\frac{r^2 B(r)}{B(r)-r^2\omega^2}\,\mathrm{d}\varphi^2.
\end{multline}
Comparing to the canonical form \cite{Rindler:2006}
\begin{equation}
\mathrm{d}s^2 = -e^{2\Phi}(\mathrm{d}t - \bar{S}_i \mathrm{d}x^i)^2+h_{ij}\mathrm{d}x^i \mathrm{d}x^j,
\end{equation}
where $x^i=(r,\varphi)$, we infer
\begin{eqnarray}\label{eq:spinComponents}
&& \Phi = \frac{1}{2} \ln\left(
B(r)-r^2\omega^2
\right),\label{eq:spinComponents1}\\
&&\bar{S}_1=0,\label{eq:spinComponents2}\\
&&\bar{S}_2 = \frac{r^2\omega}{B(r)-r^2\omega^2}, \label{eq:spinComponents3}\\
&& h_{11} = \frac{1}{B(r)},\label{eq:spinComponents4}\\
&& h_{22} = \frac{r^2 B(r)}{B(r)-r^2 \omega^2}.\label{eq:spinComponents5}
\end{eqnarray}
We assume that all the possible non-gravitational forces acting on the gyroscope are applied at its center of mass, so no torques are available in its rotating rest frame. In this regard, the spin vector $\bar{\bm{S}}$ is Fermi-Walker transported along the gyroscope's world-line. Furthermore, if we consider the orbits are on a circle of constant radius $r_g$, then it is inferred that
\begin{equation}\label{eq:Psir0}
    \left.\frac{\partial\Phi}{\partial r}\right|_{r=r_g} = 0~~\Longrightarrow~~\omega_g^2 = \frac{Q^2}{4 r_g^4}-\frac{1}{\lambda^2}.
\end{equation}
This also indicates that the curve $r=r_g$ is a geodesic and the gyroscope is indeed free falling. The above angular velocity is essentially the Kepler frequency of the orbits. The corresponding rotational rate of the gyroscope in its rest frame is given by \cite{Schiff:1960,Misner:1973}
\begin{equation}\label{eq:rotationalRate}
\Omega^2=\frac{e^{2\Phi}}{8}h^{ik}h^{jl}
\left[
\left(
\frac{\partial\bar{S}_i}{\partial x^j} - \frac{\partial\bar{S}_j}{\partial x^i}
\right)
\left(
\frac{\partial\bar{S}_k}{\partial x^l}
-\frac{\partial\bar{S}_l}{\partial x^k}
\right)
\right],
\end{equation}
which is calculated at $r=r_g$. Therefore, applying Eqs.~\eqref{eq:spinComponents1}--\eqref{eq:spinComponents5} in Eq.~\eqref{eq:rotationalRate}, we obtain
\begin{equation}\label{eq:rotationalRate-w}
    \Omega_g = \omega_g,
\end{equation}
as the rotational rate of a gyroscope orbiting in the gravitational field of a charged Weyl black hole. The gyroscope is at rest in its proper frame, however, a distant observer will detect a time dilation, which according to Eq.~\eqref{eq:rotatingFrame} is characterized by
\begin{equation}\label{eq:timeDilation_1}
    \Delta\tau = \left(
    B(r_g)-r_g^2\omega_g^2
    \right)^\frac{1}{2}\Delta t = \left(
    1-\frac{Q^2}{2r_g^2}
    \right)^\frac{1}{2}\Delta t.
\end{equation}
After a complete revolution, the orientation of the gyroscope's spin vector, relative to its rest frame, is changed by the angle 
\begin{equation}\label{eq:lag_1}
    \hat\alpha_\mathrm{rev} = \Omega_g \Delta\tau_\mathrm{rev} = \Omega_g\left(
    1-\frac{Q^2}{2r_g^2}
    \right)^\frac{1}{2}\Delta t_\mathrm{rev},
\end{equation}
where $\Delta t_\mathrm{rev} = 2\pi/\omega_g$ is the coordinate time measured in one revolution. Hence, the observed precession in the course of one orbit is calculated as $\hat\alpha'_{\mathrm{rev}} =  2\pi-\hat\alpha_\mathrm{rev}$, that by exploiting Eqs.~\eqref{eq:rotationalRate-w} and \eqref{eq:lag_1} yields
\begin{equation}\label{eq:lag_2}
    \hat\alpha'_{\mathrm{rev}} = 2\pi\left[
    1-{\left(1-\frac{Q^2}{2r_g^2}\right)^{\frac{1}{2}}}
    \right].
\end{equation}
In the case that $r_g\gg Q$, to the first order of approximation, the precession in Eq.~\eqref{eq:lag_2} becomes
\begin{equation}\label{eq:lag_3}
\hat\alpha'_{\mathrm{rev}} \approx \frac{\pi Q^2}{2 r_g^2}\quad\left(\frac{\mathrm{rad}}{\mathrm{rev}}\right),
\end{equation}
where "rad" and "rev" stand for "radians" and "revolution". The above relation has been obtained in geometric units. The value of $\hat\alpha'_{\mathrm{rev}}$ is however dimensionless and can be used to compare with the general relativistic results within proper conditions. 

The general relativistic precession for a gyroscope rotating a mass $\tilde m$ in a circular orbit of radius $r_g$, is given by \cite{Rindler:2006}  
\begin{equation}\label{eq:laq_gen}
  \hat\alpha'_{\mathrm{rev(gen)}} \approx \frac{3\pi \tilde{m}}{r_g}\quad\left(\frac{\mathrm{rad}}{\mathrm{rev}}\right)
  \end{equation}
in geometric units (for a guide to the change of units see appendix \ref{app:E}). The period of the gyroscope's orbit is easily obtained as 
\begin{equation}\label{eq:period_earth}
    \tilde{T}_{\mathrm{rev(gen)}} = 2\pi\sqrt{\frac{r_g^3}{\tilde{m}}}\quad\left(\frac{\mathrm{m}}{\mathrm{rev}}\right).
\end{equation}
Hence, using Eq.~\eqref{eq:laq_gen} and \eqref{eq:period_earth} we have
\begin{equation}\label{eq:laq_gen_1}
    \hat\alpha'_{\mathrm{rev(gen)}} \approx \frac{3 \tilde{m}^\frac{3}{2}}{2 r_g^{\frac{5}{2}}}\quad\left(\frac{\mathrm{rad}}{\mathrm{m}}\right).
\end{equation}
For the earth of mass $\tilde{m}_e \approx 4.43\times 10^{-3}\,\,\mathrm{m}$, and radius $R_e = 6371\times10^{3}\,\,\mathrm{m}$ \cite{IAU2009:2011}, if we let $r_g = R_e$, then $\tilde{T}_{\mathrm{rev(gen)}} \approx 1.52\times 10^{12}\,\,\mathrm{m}$, and the gyroscope will have approximately $6.22\times 10^3$ orbits around the earth in one year. Using the above values in Eq.~\eqref{eq:laq_gen_1} gives $\hat\alpha'_{\mathrm{rev(gen)}} \approx 4.32\times 10^{21}\,\,\mathrm{rad/m}$ $\approx 8.41$\,\,{arcsec}\footnote{1\,\,rad $=$ 206265\,\,arcsec.}$/$yr (see appendix \ref{app:E}). In the Gravity Probe B (GP-B) mission, a satellite containing four gyroscopes, was set to orbit around the earth at the altitude $r_h = 642\,\,\mathrm{km}$. The general relativistic prediction of the geodetic precession in the gyroscopic spin is therefore obtained by considering this altitude, giving 
\begin{equation}\label{eq:lag_gen_2}
\hat\alpha'_{\mathrm{rev(gen)}} \approx 8.41\left[
\frac{R_e}{R_e+r_h}
\right]^{\frac{5}{2}} \approx  6.62\quad\left(\mathrm{\frac{arcsec}{yr}}\right),
\end{equation}
which is equal 6620 mas\footnote{"mas" is an abbreviation for milliarcsec, and 1 mas = $4.848\times10^{-9}$ rad.}/yr. This value is confirmed by the reported value, $6602\pm18$ mas/yr, from the GP-B mission in 2011 \cite{GPB:2011,Everitt:2015}.

Turning back to the problem of an orbiting gyroscope around a charged source in Weyl gravity, it is plausible to adopt $r_g\equiv r_U$, where $r_U$ is the radius of circular orbits, discussed in Subsec.~\ref{subsec:circular} and derived in Eqs.~\eqref{co1}--\eqref{co3}. Accordingly, the period of the orbits, measured by a distant observer, is that given in Eq.~\eqref{eq:periodCoordinate}. If we apply these to the precession in Eq.~\eqref{eq:lag_3}, and re-scale the result, we get
\begin{equation}\label{eq:lag_4}
\hat\alpha'_{\mathrm{rev}} \approx  \left(
   1.95\times 10^{24}
   \right)\frac{Q^2 b_U}{4r_U^4}\left|
B(r_U)
\right|\quad\left(\frac{\mathrm{mas}}{\mathrm{yr}}\right),
\end{equation}
in which the numerical factor is inferred from the earlier notes in the general relativistic case and the explanations given in appendix \ref{app:E}. In this relation, as introduced before, $b_U$ is the impact parameter associated with the circular trajectories. Exploiting Eq.~\eqref{eq:LU} and the fact that $E_U^2 = V(r_U)$, yields
\begin{equation}\label{eq:bU}
    b_U \equiv \frac{L_U}{E_U} = \left|
    \frac{\omega_U}{\omega^2_U+\frac{2}{\lambda^2}-\frac{1  }{r_U^2}}
    \right|
    \quad\left(\mathrm{m}\right),
\end{equation}
where we have defined
\begin{equation}\label{eq:omegaU}
    \omega_U^2 = \frac{Q^2}{4r_U^4}-\frac{1}{\lambda^2}\quad\left(\frac{\mathrm{1}}{\mathrm{m}^2}\right).
\end{equation}
To apply a numerical assessment of $\hat\alpha'_{\mathrm{rev}}$, we need a spherically symmetric gravitating system with positive net charge. For this reason, we use the presented data in Ref.~\cite{Carvalho:2018}, where the authors have considered stability of charged white dwarfs with masses comparable to that of the sun ($M_\odot$). To elaborate this, let us consider the gyroscope is rotating such a white dwarf in a circular orbit of radius $r_U$, given in Eq.~\eqref{co3}. In Table~\ref{table:data_W}, the physical properties of the massive sources have been given. There, we have also presented the calculated values of the precession in Eq.~\eqref{eq:lag_4} for each case. Note that, the central density $\tilde{\rho}_w$ has been considered in identifying the parameter $\lambda$ of the spacetime lapse function and the value of $c_1$ has been specified in accordance to Ref.~\cite{Payandeh:2012mj} (see appendix \ref{app:E} for more details). As it is expected from Eq.~\eqref{eq:lag_4}, the precession vanishes for $Q = Q_w = 0$. Adopting a very small angular momentum (of order $\sim 10^{-8}\,\,\mathrm{m}$), we can see rather large precessions when $Q_w\neq0$. 

It is of worth to, once again, discuss the general relativistic approach. To do that however, we need to consider static charged sources whose exterior geometry is given by the Reissner-Nordstr\"{o}m (RN) metric with the lapse function \cite{Ryder:2009}
\begin{equation}\label{eq:RN_metric}
    B_{\mathrm{RN}}(r) = 1-\frac{2 \tilde m}{r}+\frac{Q_0^2}{r^2},
\end{equation}
describing spherically symmetric sources with charge $Q_0$. As mentioned before, the transition between the charged Weyl and the general relativistic geometries is not trivial. Hence, we pursue the same method, as introduced earlier in this section, to obtain the general relativistic precession in the context of charged sources. Accordingly, one obtains
\begin{equation}\label{eq:lag_RN}
    \hat\alpha'_{\mathrm{rev(RN)}} \approx \left( 1.95\times 10^{24}\right)\left[
    \frac{\sqrt{\tilde{m}}\left(
    3 \tilde{m}  r_g-Q_0^2
    \right)}{2 r_g^\frac{7}{2}}
    \right] \quad\left(\frac{\mathrm{mas}}{\mathrm{yr}}\right),
\end{equation}
assuming that $r_g\gg\tilde{m}$ and $r_g\gg Q_0$. Supposing that the gyroscope is orbiting at the altitude $r_h = 642\,\,\mathrm{km}$  around the same white dwarfs of the previous case, then $r_g = R_w + r_h$. Taking into account $\tilde{m} = M_w$, $Q_0 = Q_w$ and $M_\odot = 1.48\times10^{3}\,\,\mathrm{m}$ \cite{Sunmass:2013}, the calculated general relativistic precessions have been given in the last column of Table \ref{table:data_W}. One can observe a remarkable conformity with the results inferred from Weyl gravity for the case of $Q_w \neq 0$.\\

In this section, we assessed the dynamics of the spacetime under consideration, through a classical test of general relativity. The results are however much larger than those obtained from GP-B. This stems from the large density of macroscopic charged objects, around which we tested both the Weyl and the RN geometries. The gravitational effect of net electric charge in astrophysical objects is more significant. In the next section, we summarize the results of this paper.

\begin{table*}[t]
    \centering
        \begin{tabular}{|c|| c | c | c | c | c|}
        \hline
        $M_w/M_\odot$ & $R_w$\,\,$(\times 10^3\,\, \mathrm{m})$& $\tilde{\rho}_w$\,\,$(\times 10^{-14}\,\,\mathrm{m}^{-2})$& $Q_w$\,\,($\mathrm{m})$ & $\hat\alpha'_{\mathrm{rev}}\,\,\left(\mathrm{mas/yr}\right)$ & $\hat\alpha'_{\mathrm{rev(RN)}}\,\,\left(\mathrm{mas/yr}\right)$\\
       \hline\hline
        1.416 & 1021& 1.71316& 0& 0& $7.86833\times 10^{13}$\\
        \hline
        1.532 & 1299& 2.25971& 349.676& $1.02422\times 10^{13}$& $5.48841\times 10^{13}$\\
        \hline
        1.698 & 1539& 2.5664& 699.267& $2.32618\times 10^{13}$& $4.49524\times 10^{13}$\\
        \hline
        1.928 & 1336& 4.91076& 1053.77& $6.70766\times 10^{13}$& $6.69618\times 10^{13}$\\
        \hline
        2.203 & 1166& 14.4211& 1411.64& $2.63875\times 10^{14}$& $1.01421\times 10^{14}$\\
        \hline
        2.203 & 916.8& 29.7037& 1774.68& $6.83293\times 10^{14}$& $1.79096\times 10^{14}$\\
        \hline
    \end{tabular}
    \caption{The properties of the charged white dwarfs from Ref.~\cite{Carvalho:2018} (given in geometric units) and the values of precessions inferred from Eqs.~\eqref{eq:lag_4} and \eqref{eq:lag_RN}. For the case of precession in Weyl gravity, we have let $L=10^{-7.6}\,\,\mathrm{m}$, and the radius of orbits for the gyroscopes in the RN geometry has been put $r_g = R_w + r_h$ for each of the cases.}
    \label{table:data_W}
\end{table*}

\section{Summary and conclusions}\label{sec:conclusion}

Motion of massive particles in strongly gravitating systems is, on its own, an interesting topic in relativistic studies. In fact, such particles can indicate how these systems can construct their surroundings. Regrading the astrophysical phenomena, the particles which constitute the interstellar materials (like gases and remnants), if captured in the effective gravitational potential of large massive sources, will pursue several types of motion towards them. This, if generalized to  numerous objects, leads to the formation of stellar structures and planetary systems. Same holds for systems which include a black hole at their center. In fact, lots of galactic structures are results of the presence of a super-massive black hole at their center and the study of the orbiting objects around them, requires enough knowledge on the particle dynamics in the exterior geometry of these celestial masses. This therefore, highlights the advantage of the study of the time-like geodesics on which the particles travel at the vicinity of heavy celestial objects. 
In this paper, we paid attention to the dynamics of massive particles that approach a static black hole with net electric charge, whose exterior geometry has been inferred from the Weyl theory of gravity. We argued that the effective potential generated by this black hole can make the particles to be deflected or captured by the black hole. According to the effective potential, no planetary orbits were possible, however, the particles could be confined to an unstable circular orbit, if the particles gain specific conditions regarding their constants of motion. Particularly, we discussed the deflecting trajectories and formulated the scattering process and scrutinized it in terms of its proper cross-section. This process were further discussed for particles bounded to purely radial orbits and we indicated that this kind of motion allows for the so-called frozen particles, observed by distant observers. In the last section, we paid attention to a classical test, namely the geodetic precession, which we used to asses the effects of the under-study background geometry on the spin orbiting gyroscope. To do this, after obtaining the reliable mathematical relations, we employed a set of charged white dwarfs, as the test models. We also used the same sources to obtain the general relativistic limit of the precession, and the results indicated a good conformity between the two models. 

In conclusion, we mention that, despite the success of general relativity in passing observational tests, it still seems fruitful to pay attention to alternative theories. In the case we studied here, the black hole under consideration could generate some reliable results. So, continuing studies on alternative theories, may help us to finally overcome the remaining unsolved problems regarding the description of gravitating systems.

\begin{acknowledgements}
	M. Fathi has been supported by the Agencia Nacional de Investigaci\'{o}n y Desarrollo (ANID) through DOCTORADO Grant No. 2019-21190382. {J.R.V. is partially supported by Centro de Astrof\'isica de Valpara\'iso (CAV).}
\end{acknowledgements}

\appendix

\section{The method of solving the quartic equation $x^4-a\,x^2+b=0$}\label{app:Ap}
We are interest in solving a quartic equations of the form
\begin{equation}\label{a1}
x^4-a\,x^2+b=0,
\end{equation}
where $(a,b) > 0$ and $2\sqrt{b}\leq a$. For this purpose, we make the change of variable $x = Z \sin \vartheta$, and multiply both sides of the equation by a scalar $\alpha$. This yields
\begin{equation}\label{a2}
\alpha\, Z^4 \sin^4\vartheta - \alpha \,a\, Z^2 \sin^2 \vartheta + \alpha \,b = 0.
\end{equation}
Considering the trigonometric identity
\begin{equation}\label{a3}
4 \sin^4\vartheta -4 \sin^2 \vartheta +  \sin^2 (2 \vartheta) = 0,
\end{equation}
and comparing Eqs.~\eqref{a2} and \eqref{a3}, we infer
\begin{equation}\label{a4}
\alpha\, Z^4 =4, \qquad \alpha \,a\, Z^2=4, \qquad  \alpha \,b=\sin^2 (2 \vartheta).
\end{equation}
Solving the above equation for $Z$ and $ \vartheta$, we obtain
\begin{equation}
Z=\sqrt{a}, \quad \text{and} \quad \vartheta_n={1\over 2} \arcsin\left( {2\sqrt{b}\over a}\right) +{n\pi\over 2} ,
\label{a4}
\end{equation}
where the period of the trigonometric function is $n\pi$. Therefore, the roots of Eq.~\eqref{a3} are obtained by replacing $n = 0, 1$, giving
\begin{eqnarray}
 x_0&=&\sqrt{a}\sin\left( {1\over 2} \arcsin\left( {2\sqrt{b}\over a}\right)  \right),  \\
 x_1&=&\sqrt{a}\sin\left( {1\over 2} \arcsin\left( {2\sqrt{b}\over a}\right) +{\pi\over 2} \right) \nonumber\\
 &=& \sqrt{a}\cos\left( {1\over 2} \arcsin\left( {2\sqrt{b}\over a}\right)\right),    \\
  x_2&=&- x_0,\\
  x_3&=&-x_1.
\end{eqnarray}
The above method enables us to determine the black hole horizons.

\section{Solving Eq.~\eqref{eq:Vprime=0} using the Cardano's method}\label{app:A}

Equation \eqref{eq:Vprime=0} can be reduced into an equation of the third order, by applying the change of variable $X\doteq r^2$. Accordingly, the reduced equation becomes
\begin{equation}\label{eq:Vprime-reduced}
    4 X^3 + a_1 X - a_2 = 0,
\end{equation}
in which we have used
\begin{eqnarray}\label{eq:a12}
    a_1 &=& 4\lambda^2\left(L^2-\frac{Q^2}{4}\right),\\
    a_2 &=& 2 \lambda^2 L^2 Q^2.
\end{eqnarray}
For $a_1 = 0$ (i.e. $L = Q/2$), the equation is easily solved as $X^3 = a_2/4$ and we get the value in Eq.~\eqref{co2}. Since always $a_2 > 0$, the general form of the equation only varies depending on the sign of $a_1$. Accordingly, we compare Eq.~\eqref{eq:Vprime-reduced} by two hyperbolic identities
\begin{eqnarray}
    4 \sinh^3\vartheta + 3\sinh\vartheta -\sinh(3\vartheta) &=&0,\label{eq:identity1}\\
     4 \cosh^3\vartheta - 3\cosh\vartheta -\cosh(3\vartheta) &=&0.\label{eq:identity2}
\end{eqnarray}
The following two cases are available:
\begin{itemize}
    \item For $L>Q/2$: Since $(a_1, a_2) >0$, then defining $X \doteq \Xi_0 \sinh\vartheta$, we recast Eq.~\eqref{eq:Vprime-reduced} as
    \begin{equation}\label{eq:Vprime_reduced_i1}
        \ell\, \Xi_0^3\sinh^3\vartheta+a_1 \ell\, \Xi_0 \sinh\vartheta-a_2\, \ell = 0,
    \end{equation}
   in which, $\ell$ is a Legendre coefficient. Comparing Eqs.~\eqref{eq:Vprime_reduced_i1} and \eqref{eq:identity1}, we get
    \begin{eqnarray}\label{eq:values}
        &&\ell = \frac{4}{\Xi_0^3},\label{eq:values1}\\
       && \Xi_0 = \sqrt{\frac{4 a_1}{3}},\label{eq:values2}\\
       && \sinh(3\vartheta) = \sqrt{\frac{27 a_2^2}{4 a_1^3}}\doteq\Xi_1.\label{eq:values3}
    \end{eqnarray}
    It is therefore inferred that $\vartheta = \frac{1}{3}\sinh^{-1}\Xi_1$, resulting in 
    \begin{equation}\label{eq:X1}
        X  = \Xi_0 \sinh\left(
        \frac{1}{3} \arcsinh(\Xi_1)
        \right),
    \end{equation}
   and the value in Eq.~\eqref{co1} is followed for the unstable orbits in the case of $L>Q/2$. 
   
   \item For $L<Q/2$: This time, since $a_1<0$ and $a_2>0$, the comparison is made to Eq.~\eqref{eq:identity2}, by means of the definition $X \doteq \Xi_0 \cosh\vartheta$. Pursuing the  same procedure as the previous case, we obtain 
   \begin{equation}\label{eq:X2}
        X  = \Xi_0 \cosh\left(
        \frac{1}{3} \arccosh(\Xi_1)
        \right),
    \end{equation}
    and we get the radius in Eq.~\eqref{co3}.
    
\end{itemize}

\section{The method of obtaining $r_P$ and $r_A$}\label{app:B}

The method is similar to that used in appendix \ref{app:A}. The equation $\mathfrak{P}(r) = 0$ produces
\begin{equation}\label{eq:p(r)=0}
   X^3 - \alpha X^2 - \beta X + \gamma = 0\,\quad\quad (X=r^2),
\end{equation}
which by performing the Tschirnhaus transformation $S = X - \alpha/3$, gives 
\begin{equation}\label{eq:p(r)=0-reduced}
   S^3 - \bar{a}_1 S - \bar{a}_2 = 0,
\end{equation}
in which
\begin{eqnarray}\label{eq:new-a1,a2}
    \bar{a}_1 &=& \frac{4}{3}\left(
    \alpha^2+3\beta
    \right),\\
    \bar{a}_2 &=& 4\left(
    \frac{2\alpha^3}{27}+\frac{\alpha\beta}{3}-\gamma
    \right).
\end{eqnarray}
Considering the trigonometric identity
\begin{equation}\label{eq:identity_3}
    4\cos^3\vartheta - 3 \cos\vartheta - \cos(3\vartheta) = 0,
\end{equation}
we define $S = \xi_0 \cos\vartheta$ and recast Eq.~\eqref{eq:p(r)=0-reduced} as
\begin{equation}\label{eq:P(r)_reduced_1}
    \ell\, \xi_0^3 \cos^3\vartheta - \ell\, \bar{a}_1 \xi_0 \cos\vartheta - \ell\, \bar{a}_2 = 0.
\end{equation}
As in the previous cases, comparing the above equations, we obtain
\begin{eqnarray}\label{eq:xi0,xi1}
    \xi_0 &=& 2\sqrt{\frac{\beta}{3}+\frac{\alpha^2}{9}},\\
  \xi_1 &=&  \left(\frac{8\alpha^3}{9}+4\alpha\beta-12\gamma\right)\sqrt{\frac{3}{(4\beta+\frac{4\alpha^2}{3})^3}},
\end{eqnarray}
where $2 n\pi$ indicates the periodic symmetry of the cosine function. Accordingly, and using the reverse transformations, the solutions to $\mathfrak{P}(r)$ can be given as
\begin{equation}\label{eq:P(r)=0_new_solutions}
    r_n = \left[
    \xi_0 \cos\left(\frac{1}{3}\arccos(\xi_1)+\frac{2 n\pi}{3}\right)
    +\frac{\alpha}{3}
    \right]^\frac{1}{2}.
\end{equation}
The above solution results in positive values for $n=0,2$ and is periodically repeated as $n\rightarrow n+3$. We can therefore take two different values as physically meaningful solutions to our equation, by designating $r_A = r_{n=0}$ and $r_P = r_{n=2}$ which is in agreement with $r_A>r_P$.

\section{Solving the angular equation of motion}\label{app:C}

The change of variables applied in solving $\mathfrak{P}(r)=0$ can not make a simple reduction of order to solve the differential equation in Eq.~\eqref{tl11}. In fact, this kind of definition provides a fourth order elliptic integral equation which, although doable, is hard to solve. We therefore propose a more efficient method for this particular case (for a good introduction to the methods of reducing fourth order elliptic integrals well-defined solutions, see Ref.~\cite{prasolov1997elliptic}). Since the scattering happens at $r_P$, we instead, define the following non-linear change of variable:
\begin{equation}\label{eq:c-x}
    x \doteq \left(\frac{r_A}{r}\right)^2,
\end{equation}
producing $\mathrm{d}r=-r_A\left(
\frac{\mathrm{d}x}{2x^{3/2}}
\right)$ which reduces Eq.~\eqref{tl11} to
\begin{equation}\label{eq:c_reducedIntegral_1}
    \mathrm{d}\phi = \pm L \lambda \frac{-r_A\,\mathrm{d}x}{2\sqrt{\gamma\,\tilde{\mathfrak{P}}(x)}},
\end{equation}
in which
\begin{equation}\label{eq:c-reduced_1}
 \gamma\,\tilde{\mathfrak{P}}(x) \equiv x^3\mathfrak{P}(x) = \gamma\left(x^3 - \tilde\alpha x^2 - \tilde\beta x + \tilde\gamma
 \right),
\end{equation}
where
\begin{eqnarray}\label{eq:c_alpabetagammabar}
 && \tilde{\alpha} = \frac{\beta\,r_A^2}{\gamma},\\
 && \tilde{\beta}=\frac{\alpha\,r_A^4}{\gamma},\\ &&\tilde{\gamma} = \frac{r_A}{\gamma}.
    \end{eqnarray}
A further change of variable 
\begin{equation}\label{eq:c-u}
    u\doteq\frac{1}{4}\left(
    x-\frac{\tilde{\alpha}}{3}
    \right),
\end{equation}
leads to the following reduced integral form of Eq.~\eqref{eq:c_reducedIntegral_1}:
\begin{equation}\label{eq:c-reduced-Weierstrass}
\int_{\phi_0}^\phi \mathrm{d}\phi' = \pm\frac{2\sqrt{\gamma}}{L \lambda\, r_A}\int_{u_A}^u \frac{-\mathrm{d}u'}{\sqrt{\mathfrak{P}(u')}},
\end{equation}
in which $u_A = \frac{1}{4}(1-\frac{\beta \,r_A^2}{3\gamma})$, and 
\begin{equation}\label{eq:c-P(u)}
    \mathfrak{P}(u) =4 u^3-g_2 u -g_3, 
\end{equation}
where 
\begin{eqnarray}\label{eq:c-g2g3}
    g_2 &=& \frac{1}{4}\left(\frac{\tilde{\alpha}}{3}+\tilde{\beta}\right),\\
    g_3 &=& \frac{1}{16}\left(
    \frac{2\tilde\alpha^3}{27}+\frac{\tilde{\alpha}\,\tilde{\beta}}{3}
    -\tilde{\gamma}
    \right), 
\end{eqnarray}
are the Weierstra$\ss$ coefficients, associated with the third order polynomial $\mathfrak{P}(u)$. Recasting Eq.~\eqref{eq:c-reduced-Weierstrass}, we have
\begin{multline}\label{eq:c-reduced-Weierstrass-1}
    \pm\frac{2\sqrt{\gamma}}{L\lambda\,r_A}(\phi-\phi_0) 
    = -\left\{
    \int_{u_A}^\infty \frac{\mathrm{d}u'}{\sqrt{\mathfrak{P}(u')}}
    - \int_u^\infty\frac{\mathrm{d}u'}{\sqrt{\mathfrak{P}(u')}}
    \right\}\\
    = -\left\{
    \ss(u_A) - \ss(u)
    \right\},
\end{multline}
in which have used the definition 
\begin{equation}\label{eq:c-WeierstrassDef}
    \ss(u)\equiv\wp^{-1}\left(u;g_2,g_3\right)
    = \int_u^\infty \frac{\mathrm{d}u'}{\sqrt{4 u'^3-g_2 u'-g_3}}
\end{equation}
of the inverse $\wp$-Weierstra$\ss$ function \cite{handbookElliptic}. Accordingly, using the values of $\tilde{\gamma}$ and $\gamma$, and defining $\varphi_A = \ss(u_A)$, from Eq.~\eqref{eq:c-reduced-Weierstrass-1} we deduce
\begin{equation}\label{eq:c-u_1}
    u(\phi) = \frac{1}{4}\left(
    \frac{r_A^2}{r^2(\phi)}
    -\frac{\beta\,r_A^2}{3\gamma}
    \right)
    = \wp\left(\pm\frac{2\sqrt{\gamma}}{L \lambda\,r_A}(\phi_0-\phi)
    +\varphi_A
    \right),
\end{equation}
which for $\phi_0 = 0$ results in the solution
\begin{equation}\label{eq:c-r_1}
r(\phi) = \frac{r_A}{\sqrt{4 \wp\left(\varphi_A\mp\frac{2\sqrt{\gamma}}{L \lambda\,r_A}\phi\right)+\frac{\beta\,r_A^2}{3\gamma}}}.
\end{equation}

\section{Switching the values of spacetime coefficients and dynamical quantities between the geometric and SI units}\label{app:E}

The values of the Newton's gravitational constant and the speed of light are \cite{CODATAG:2009,CODATAc:2009}
\begin{eqnarray}\label{eq:Gc}
    &&G = 6.67430 \times 10^{-11}\quad (\mathrm{m^3 {kg^{-1}} s^{-2}}),\\
    &&c = 299792458\quad (\mathrm{m\,s^{-1}}).
    \end{eqnarray}
The mass of earth is $m_e = 5.97237 \times 10^{24}\,\,\mathrm{kg}$ \cite{IAU2009:2011} which in geometric units becomes
\begin{equation}\label{eq:mbe}
    \tilde{m}_e = m_e\times \frac{G}{c^2} = 4.4347\times 10^3\quad (\mathrm{m}).
\end{equation}
In the geometric units, the value of time is also given in meters by applying $\bar\tau\,\, (\mathrm{m})= \tau\times c\,\,(\mathrm{s})$. For example, one year is about $3.1536\times10^7\,\,\mathrm{s}$, which in meters is equivalent to $1\,\mathrm{yr} = 9.45\times 10^{15}\,\,\mathrm{m}$. 

The change of units from Coulomb (C) to meters for the electric charge $Q$, is also done as bellow:
\begin{equation}\label{eq:QcQm}
 \left[ Q\,\,\mathrm{(m)}\right] = 
    \left[Q\,\,(\mathrm{C})\right]\times\sqrt{\frac{G}{4\pi\varepsilon_0c^4}},
\end{equation}
in which $\varepsilon_0 = 8.854 \times 10^{-12}\,\frac{\mathrm{C}^2}{\mathrm{N} \mathrm{m}^2}$ is the vacuum permittivity \cite{CODATAepsilon:2009}. This way,
\begin{equation}\label{eq:QcQm_1}
      [Q\,\,(\mathrm{C})] = \left(
    1.15964\times10^{17}
    \right)  [Q\,\,\mathrm{(m)}].
\end{equation}
Furthermore, the factor $1/\lambda^2$ in Eq.~\eqref{par1} is a density of dimensions $\mathrm{m}^{-2}$. In fact $\lambda$ is given by
\begin{equation}\label{eq:lambda}
    \lambda = \left[
    3\tilde\rho_c + \frac{2}{3}c_1
    \right]^{-\frac{1}{2}}\quad(\mathrm{m}),
\end{equation}
in which $\tilde{\rho}_w$ is the density of a spherically symmetric charged massive source. Here, we let 
$c_1 = 2.08\times 10^{-54}\,\,\mathrm{m}^{-2}$, as given in Ref.~\cite{Payandeh:2012mj} and is comparable to the value of the cosmological constant $\Lambda_0 = 1.1056\times 10^{-52}\,\,\mathrm{m}^{-2}$ \cite{Planck:2015}.

In geometric units, the dimension of angular momentum is square meters, which is transformed to the SI units [$\mathrm{kg\,m^2\,s^{-1}}$] by applying a $c^3/G$ multiply. However, since we have ignored the mass of the orbiting objects, the value of the constant of motion $L$, in geometric units, is given in meters which is in conformity with the other dynamical quantities.

Taking into account the above notes and working in the geometric units, the value of precession will be the same as that in the SI units.

\bibliographystyle{spphys} 
\bibliography{Biblio_v1.bib}

\end{document}